\documentclass[prd,aps,preprint,amsmath,nofootinbib,amssymb,eqsecnum,showkeys,tightenlines]{revtex4-1}
\usepackage{slashed}
\usepackage{epsfig,latexsym,cancel,amssymb,amsmath,verbatim,mathrsfs}
\usepackage{color}
\usepackage{graphicx}
\usepackage{caption}
\usepackage{subcaption}
\usepackage{hyperref}

\definecolor{My_red}{cmyk}{0.00,1.00,1.00,0.20}

\usepackage{float}
\usepackage{tikz-feynhand}
\newcommand{\secref}[1]{Section~\ref{#1}}
\newcommand{\figref}[1]{Fig.~\ref{#1}}

\def\ra{\rightarrow}
\def\L{\left(}
\def\R{\right)}
\def\wt{\widetilde}

\def\ld{\lambda}
\def\f{\frac}
\graphicspath{{images/}}
\usepackage{multirow}

\begin{document}


\title{A Light Lepton-flavor-violating Flavon: \\
The Messenger of Neutrino Mixing and Muon $g-2$ }
\author{Shuyang Han}
\email[E-mail: ]{d201980123@hust.edu.cn}
\affiliation{School of physics, Huazhong University of Science and Technology, Wuhan 430074, China}

\author{Zhaofeng Kang}
\email[E-mail: ]{zhaofengkang@gmail.com}
\affiliation{School of physics, Huazhong University of Science and Technology, Wuhan 430074, China}



\date{\today}

\begin{abstract}

In neutrino physics, a class of models with local or global family symmetries may be invoked, and then a flavon field is needed to realize the full neutrino mixing. This flavon may shed light on the long-standing muon $g-2$ puzzle. In this work, we explore this idea in the $({B-L})_{13}$ gauge extension to the standard model (SM), in which realistic neutrino mixing requires both a SM singlet flavon $s$ and a vector-like lepton (VLL) doublet. The dominant coupling between the flavon and leptons is in the manner of lepton-flavor-violation (LFV). Through an analytical analysis of the SM lepton-VLL mixing matrix, we find that the parameter space of the $s\bar\mu e$-type flavon to explain the muon $g-2$ has been completely excluded by the specific LFV process, muonium-antimuonium oscillation. But the $s\bar\mu \tau$-type flavon still has the opportunity; however, it confronts the strong constraint from $\tau\to \mu$ conservation and, in particular, the lepton flavor universality test of $Z$ boson decay, which arises due to our way to realize the LFV flavon. The surviving flavon is highly predictable, with mass in the narrow window $m_\tau\lesssim m_s\lesssim 1.5~ m_\tau$ and LFV coupling strength $\sim 10^{-2}$. Besides, it leaves a TeV scale VLL with a multi-lepton signature at the LHC.

\end{abstract}


\maketitle

\newpage

\tableofcontents

\newpage

\section{Introduction}

Although the establishment of the Standard Model (SM) of particle physics achieves great success, it represents just a low-energy manifestation of the more fundamental theory and confronts many puzzles. We have been committed to solving these puzzles in order to unravel the mysterious mask of fundamental theories. The tiny but nonvanishing neutrino mass is a well-established puzzle, and therefore the neutrino extension furnishes a good basement on the way to searching for new physics. Of interest, there is another long-standing puzzle in the leptonic sector. That is the anomalous magnetic moment of the muon lepton. Since 2006, when Brookhaven National Laboratory (BNL) first reported the deviation of the measured muon  $g-2$ value from the SM prediction~\cite{Muong-2:2006rrc}, we have experienced many anomalies in particle physics, but none of them were confirmed. However, recently this anomaly has been confirmed by the Fermi National Accelerator Laboratory (FNAL)~\cite{Muong-2:2021ojo,Muong-2:2023cdq,Muong-2:2021vma,Muong-2:2021ovs,Muong-2:2021xzz}. Now, the discrepancy between the experimental results~\cite{Muong-2:2021ojo} and the SM prediction~\cite{Aoyama:2020ynm}, combining the two results, is 
\begin{equation}
    a_\mu^{{\rm Exp}}-a_\mu^{{\rm SM}}=(251\pm59)\times 10^{-11},
\end{equation}
with a significance of $4.2~\sigma$. Despite the QCD uncertainty caused by the lattice calculation of hadronic vacuum polarization and hadronic light-by-light scattering, this robust excess still carries the hope of many people for new physics.

It is tempting to conjecture that the muon $g-2$ puzzle may be a consequence of solving the neutrino puzzle. Put in a different way, the muon $g-2$ deviation conveys to us information about the origin of neutrino mass and, more likely, mixing, which signifies why the muon flavor is selected. This work is attempting to explore new physics along such a line. Let us start from the neutrino side, and it probably implies the elegant seesaw mechanism. This mechanism can be elegantly realized in the local $B-L$ extension to the SM~\cite{Davidson:1978pm,Mohapatra:1980qe,Marshak:1979fm}. In this model, there is a natural candidate to enhance muon $g-2$ value, the new massive gauge boson $Z'$. However, owing to coupling both to quarks and the electron, it is unable to fulfill this role by virtue of many related constraints. Such a situation motivated the flavorful local $(B-L)_{23}$ scheme~\cite{Kang:2019vng} which only puts the second and third families of fermions to be charged under the new gauge group, thus providing an electron/nucleon phobic $Z'$ at the sub-GeV scale to explain the muon $g-2$ data.

In this work, we consider another candidate, a light flavon with lepton flavor violation (LFV), to enhance muon $g-2$ in the gauged $(B-L)_{ij}$ model. To be more general, such a flavon may be present in any model with a family symmetry $G_X$. This symmetry forbids certain family mixings, and usually a flavon field ${\cal F}$ spontaneously breaking $G_X$ is indispensable; sometimes a vector-like family of fermions $F=(F_L,F_R)$ is also introduced to mediate the $G_X$ breaking effect on SM fermions $f_i$ through coupling $~{\cal F}\bar f_i F$, thus producing full mass mixing of fermions~\cite{Han:2024rwz}. In general, those couplings give rise to FCNC, and for instance,  for the $Z'$ studied in Ref.~\cite{Kang:2019vng}, the muon $g-2$ favored region may be threatened by the $s\to d ~Z'$ transition at the loop level~\cite{Kang:2020gfi}. This situation partially motivates us to carry out the current study, to explore the possibility of the light LFV flavon in solving the muon $g-2$ puzzle.

To demonstrate the idea, let us consider the local $(B-L)_{13}$ and restrict the discussion in the leptonic sector. Then, from ${\cal F}\bar \ell_2 F$ we obtain a flavon with LFV coupling to $\bar\mu e$ or $\bar\mu \tau$, depending on the mixing between light and exotic heavy charged leptons. Setting the flavon (CP-even in our work) to be as light as possible,not much heavier than $m_\mu$ and $m_\tau$ for the mentioned two scenarios, respectively, we then have the potential to solve the muon $g-2$ puzzle via the LFV flavon loop without running heavy particles. This opens up a new scenario to understand this long-standing puzzle, different from the usual scheme through a muonic boson far below the weak scale, which conserves the lepton flavor~\cite{Baek:2001kca,Pospelov:2008zw,Liu:2018xkx,Kang:2019vng,Han:2021gfu,Zhevlakov:2023jzt}. Actually, we find that this is only relatively new, and similar light bosons, axion-like and spin-1, are already present, and they appear both in the effective models and concrete flavorful models~\cite{Dev:2017ftk,Bauer:2019gfk,Cornella:2019uxs,Calibbi:2020emz,Ema:2022afm,Kriewald:2022erk}.

Of course, in a concrete model, whether the scenario of light LFV flavon can succeed requires a detailed analysis of the strong constraints of lepton flavor physics (such as charged lepton flavor decay, lepton flavor universality violation of $Z$-boson decay, muonium and anti-muonium oscillation, etc.) as well as the related constraints on the light boson. To arrive at the feasible parameter space in the $(B-L)_{13}$ model, we consider the minimal heavy-light mixing which facilitates approximate analytical analysis. We find that the flavon of $s\bar\mu e$ type has been ruled out, which confirms the conclusion for the LFV axion-like particle in Ref.~\cite{Endo:2020mev}. The $s\bar\mu \tau$-type LFV flavon still possesses the opportunity to solve the puzzle, leaving a promising TeV scale VLL with multi-lepton signature at the LHC; besides, it predicts a flavon with mass close to $m_\tau$ with LFV coupling strength $\sim 10^{-2}$.

The work is organized as follows. In Section II we present the local $(B-L)_{13}$ model for LFV flavon, and discuss the constraints from $Z$ boson decay. In Section III we investigate what kind of LFV flavon can successfully resolve the muon $g-2$ puzzle. The final section includes the conclusion and discussions.

\section{flavon \& vector-like lepton from the local \texorpdfstring{$(B-L)_{13}$}{B-L13} model}

In this work, as a concrete realization of LFV flavon for the muon $g-2$ puzzle, we work in the gauged $(B-L)_{13}$ extension to the SM, under which only the first and third generation of fermions are charged. From first glance, the leptonic sector of this model is similar to that of the well-known gauged $L_e-L_\tau$ model. However, they have a striking difference: here we have two RHNs for the sake of anomaly cancellation, and they are able to implement the minimal seesaw mechanism.   

Although successfully explaining neutrino masses, the minimal $(B-L)_{13}$ model cannot fully account for neutrino mixing, that is, the Pontecorvo-Maki-Nakagawa-Sakata (PMNS) matrix~\cite{Maki:1962mu,Pontecorvo:1967fh}, due to the flavorful $B-L$. Extra ingredients are still necessary to give realistic neutrino mixing. In this article, like the scheme in~\cite{Kang:2019vng}, we rely on the vector-like fermions $L_{L,R}$ and a flavon $\mathcal{F_\ell}$ to do this job by assigning a suitable $(B-L)_{13}$ charge. Then, the relevant field content of the model is listed in Table~\ref{tab:QN}. Given the above field content and symmetries, the most general Lagrangian of the leptonic sector reads 
\begin{align}\label{eq:LagB_L23}
-\mathcal{L}_L & = Y^e_{2 2} \bar{\ell}_{2} H e_{R 2} + Y^{e}_{i j} \bar\ell_{i} H e_{R j} + Y^{N}_{i j} \bar\ell_{i} \tilde{H} N_{Rj} 
+ Y^e_{i} \overline{L}_L H e_{Ri} + Y^N_i \overline{L}_L \widetilde{H} N_{Ri} \\
\nonumber
& \hspace{0.9em}
+\lambda_2^{\ell} \bar{\ell}_2 L_R \mathcal{F}_{\ell}^* 
+ M^{\ell}_i \bar{\ell}_i L_R + m_L \overline{L}_L L_R + \frac{\lambda^N_{i j}}{2} \Phi \bar N^{C}_{R i} N_{R j} + h.c. \, ,
\end{align}
where $\wt H = i \sigma_2 H^{*}$, the Latin indices $i/j = 1,3$. It is convenient to work in the basis where $Y^e_{i j}$ and $\ld^N_{i j}$ are diagonal. The Yukawa term $\ld_{2}^\ell\bar\ell_2L_R{\cal F}_{\ell}^*$ is crucial because it is the only term that encodes the information of flavorful $(B-L)_{13}$ spontaneously breaking by the flavon field, which is written as $\mathcal{F_\ell}=\frac{v_f}{\sqrt{2}}+\frac{s+ia}{\sqrt{2}}$, and then mediates it to the SM sector through a light-heavy doublet mass mixing, to generate full neutrino mixing  along with the $Y_i^N$-Dirac mass terms~\footnote{To discuss neutrino physics in this model, it is convenient to start from an effect minimal seesaw model after integrating out the VLLs. Then, the missed mixing elements in the seesaw sector of the original $(B-L)_{13}$ appear as the effective Dirac mass terms $Y_{1,3}^N \lambda_2^\ell v_f$. In this work, we will not elaborate on the details and believe that the free $Y_i^N$ coupligns and RHN masses ensure that it is just a matter of data fitting to reproduce the realistic neutrino mixing.}. On top of that, this Yukawa coupling makes the flavon obtain LFV coupling to $\bar \mu e$ or $\bar \mu\tau$, by virtue of the mixing between $L_R^-$ and the corresponding light charged lepton species, which opens the possibility to solve the muon $g-2$ puzzle through a light LFV flavon. 
\begin{table}[htb]
\begin{center}
\begin{tabular}{|c|c|c|c|c|c|c|c|}
\hline\hline
  & $\ell_{e},\ell_{\mu},\ell_{\tau}$ & $e_{R},\mu_{R},\tau_{R}$ &  $N_{e R},N_{\tau R}$ & $H$ & $\Phi$ & $L_{L/R}$ & $\mathcal{F}_{\ell}$  \\
\hline 
$SU(2)_L\times U(1)_Y$ & $(2,-1/2)$ & $(1,-1)$ & $(1,0)$ & $(2,1/2)$&$(1,0)$ & $(2,-1/2)$ &$ (1,0)$ \\
\hline
$(B-L)_{13}$ & $-1,0,-1$ &$ -1,0,-1$ & $-1,-1$  & 0 & 2 & $-1$ &$ -1$ \\
\hline\hline
\end{tabular}
\end{center}
 \captionsetup{justification=centering}   
\caption{Field content and quantum numbers in the $({B-L})_{13}$ model.}
\label{tab:QN}
\end{table}

We have not specified the scalar potential of the model, $V(\Phi,\mathcal{F}_\ell)$, assuming that the two new scalars almost decouple from the SM Higgs doublet $H$. We will not delve into the details and just point out its main features that are relevant to our subsequent discussions. This potential should be designed to generate both vacuum expectation values (VEV) for the two scalar fields, $\langle\mathcal{F}_\ell\rangle$ and $\langle \Phi\rangle$, 
with the latter related to RHN masses. Next, both the CP-odd and CP-even components of the flavon, $a$ and $s$, can be the required LFV flavon, but in this work we take $s$ as an example under the assumption that the effect of $a$ is minor, with a comment soon later. It may be ascribed to the fact that $a$ is the dominant Goldstone boson mode of $(B-L)_{13}$ spontaneously breaking, thus eaten by the gauge boson; otherwise, $a$ is considerably heavier than $s$ as a result of the mixing effect between the CP-even components of $\mathcal{F}_\ell$ and $\Phi$~\cite{Kang:2012sy}.

Comments are in order. First, the subsequent analysis proceeds in the limit of $g_{B-L}\to 0$, and thus the local $(B-L)_{13}$ effectively becomes a global one. This simplification frees us of any constraints associated with the new massive gauge boson. Second, as studied in Ref.~\cite{Bauer:2019gfk}, contrary to the flavor conservation pseudo scalar boson, the CP-odd component $a$ having a nondiagonal coupling to $\bar e\mu$ moreover heavier than $m_\mu$ can also make a positive contribution to muon $g-2$ at the one-loop level. Therefore, we can instead pick up $a$ as the LFV flavon to enhance muon $g-2$, but it is expected that such a switch will not give a substantial difference to the option in the current work. Last, $(B-L)_{13}$ may not be the unique option, and for instance, the alternative $(B-L)_{12}$ instead needs the Yukawa coupling $\ld_{3}^\ell\bar\ell_3L_R{\cal F}_{\ell}^*$ which also admits a LFV flavon coupling to $\bar \tau\mu$.

\subsection{The charged lepton mass mixing with the Occam razor principle} 

Contrary to the ordinary known leptons, the vector-like leptons (VLLs) transform under the non-chiral representation of SM, see~\cite{Dreiner:2008tw,Martin:2012us} for reviews. Therefore VLL can possess electroweak-singlet mass and it will provide the possibility of  large mixing~\cite{Dermisek:2013gta,Kannike:2011ng,Arkani-Hamed:2021xlp,Lee:2022nqz}. 

The mass terms for the charged leptons, including the heavy vector-like pair, are given by $(\bar{e}_{L}) {M}_e(e_{R})$ with the most general mass mixing matrix restricted by the flavorful $B-L$ symmetry taking the following form
\begin{equation}\label{Me:full}
{M}_e\equiv
\begin{pmatrix}
 \frac{v_h}{\sqrt{2}} Y^{e}_{11} & 0 & 0 & {M}^{\ell}_{1} \ra 0\\[0.5ex]
 0 & \frac{v_h}{\sqrt{2}} Y^{e}_{22} & 0 & {M}^{\ell}_{2} \\[0.5ex]
 0 & 0 & \frac{v_h}{\sqrt{2}} Y^{e}_{33} & {M}^{\ell}_{3}\ra 0 \\[0.5ex]
 \frac{v_h}{\sqrt{2}} Y_1^e & 0 & \frac{v_h}{\sqrt{2}} Y_3^e\ra 0 & {m}_L\\[0.5ex]
\end{pmatrix}
\sim
\begin{pmatrix}
 m_{e} & 0 & 0 &  0 \\[0.5ex]
 0 & m_{\mu} & 0 &  0 \\[0.5ex]
 0 & 0 & m_{\tau} & 0 \\[0.5ex]
 0 & 0 & 0 & m_{4} \\[0.5ex]
\end{pmatrix} \, ,
\end{equation}
with $m_L$ around the TeV scale, the largest mass scale in $M_e$. The Dirac mass terms between the light and heavy leptons are denoted as $M^{\ell}_{a}$ with $a = 1,2,3$, and especially the element 
$M^{\ell}_{2}=\lambda_2^{\ell}\frac{v_f}{\sqrt{2}}\neq 0$ is a result of the flavorful $(B-L)_{13}$ spontaneously breaking by the flavon field, which is indispensable to realize full neutrino mixing. Through bi-unitary transformation $[e_{L,R}]_{a} \rightarrow [\hat{e}_{L,R}]_{a} = [U_{L,R}^{\dagger}]_{a \alpha} [e_{L,R}]_{\alpha}$, we move to the mass basis of charged leptons: ${M}_e
=U_{L}
\hat{\mathcal{M}}_e 
U_{R}^{\dagger}$ with $\hat{\mathcal{M}}_e ={\rm diag}(m_e,m_\mu,m_\tau,m_4)$. Hereafter, we will drop the hat index for the states in the mass eigenstates.

Be careful that $M_e$ contains several non-diagonal elements that have the potential to cause a large LFV. Let us use the Occam's razor principle to control the number of relevant parameters. From the side of neutrino physics, $M_{1,3}^\ell$ can be safely sent to zero, and $Y_{1,3}^e$ is also irrelevant. However, we will find that at least one of them is crucial to provide a feasible solution to the muon $g-2$ puzzle through the LFV coupling of the light flavon. For example, let us focus on the case with $Y_1^e\neq0$. Then, the mixing effects of the charged lepton mass matrix $M_e$ are encoded in the following $3\times 3$ matrix with five elements, 
\begin{equation}\label{3by3}
\begin{split}
M =
\begin{pmatrix}
    Y^{e}_{11}\frac{v}{\sqrt{2}} & 0 & 0\\[0.5ex]
    0 & Y^{e}_{22}\frac{v}{\sqrt{2}} & M^{\ell}_{2}\\[0.5ex]
    Y^{e}_{1}\frac{v}{\sqrt{2}} & 0 & m_L\\[0.5ex]
\end{pmatrix} 
\equiv  m_L
\begin{pmatrix}
    y_e & 0 & 0  \\[0.5ex]
    0 & y_\mu & a  \\[0.5ex]
    b & 0 & 1 \\[0.5ex]
\end{pmatrix}. \, 
\end{split}
\end{equation}
The mixing elements in $U_L$ and $U_R$ are obtained by diagonalizing the Hermitian  matrices $M M^{\dagger}/|m_L^2|$ and $M^{\dagger} M/|m_L^2|$, respectively  
\begin{equation}\label{msq}
\begin{split}
\frac{M M^{\dagger} }{\left|m_L\right|^2}
= 
\begin{pmatrix}
    \left|y_e\right|^2 & 0 & y_e b^*
    \\[0.5ex]
    0 & \left|y_\mu\right|^2 + \left|a\right|^2 & a 
    \\[0.5ex]
    y_e^* b & a^* & \left|b\right|^2 + 1 
    \\[0.5ex]
\end{pmatrix} \,,~~~~
\frac{M^{\dagger} M }{\left|m_L\right|^2} 
=  
\begin{pmatrix}
    \left|y_e\right|^2 +\left|b\right|^2 & 0 & b^*
    \\[0.5ex]
    0 & \left|y_\mu\right|^2 & y_\mu^* a 
    \\[0.5ex]
    b & y_\mu a^* & \left|a\right|^2 + 1 
    \\[0.5ex]
\end{pmatrix} \,.
\end{split}
\end{equation}
It is seen that $a$ and $b$ can generate 4-2 and 4-1 mixing in $U_L$, respectively, but the latter is suppressed by the chiral flip from the first generation, $y_e$. After $a\leftrightarrow b$ and $y_e\leftrightarrow y_\mu$, similar behaviors are obtained in $U_R$.

In our model, the performances of $U_L$ and $U_R$ in the LFV processes are quite different. For the moment, let us consider their imprints within the SM content, but we do not take into account the charged current, since it involves the unknown neutrino sector. Then, first, in the neutral current of the $Z$ boson, $U_L$ is gone after moving to the mass basis due to the family universe, including $L_L$. Next, in terms of the discussion in Appendix.~\ref{h-LFV}, the LFV related to the SM Higgs boson can be written as 
\begin{equation}\label{eq:HiggsLFV}
    \mathcal{L}_{h}=-\hat Y^e_{ij}h\bar e_i P_R e_j+h.c. \quad {\rm with}\quad \hat{Y}^{e}_{i\neq j}=-\left(U_L^\dagger\frac{M_e(v_h\to 0)}{v_h} U_R\right)_{i\neq j}.
\end{equation}
Thus, for the ansatze like Eq.\eqref{3by3}, $\hat{Y}^{e}_{i\neq j}=-\frac{m_L}{v_h}(U_L^*)_{4i}(U_R)_{4j}-\frac{v_f}{v_h} {Y}^{s}_{i\neq j}$ 
with $Y^s$ the Yukawa coupling matrix of the flavon given in Eq.\eqref{flavon:LFV}. These non-diagonal elements lead to radiative charged lepton flavour violating (cLFV) such as $\mu \to e\gamma$ and $\tau \to \mu \gamma$, thus being severely constrained. From~\cite{Harnik:2012pb}, we get the upper limits $\sqrt{|\hat{Y}^{e}_{2 1}|^2+|\hat{Y}^{e}_{2 1}|^2}<3.6\times10^{-6}$ and $\sqrt{|\hat{Y}^{e}_{23}|^2+|\hat{Y}^{e}_{3 2}|^2}<0.016$. They are not of concern since the similar constraints from the light LFV flavon, which will be studied later, are much stronger. In the following subsection, we will show that within SM,  the decay of the $Z$ boson is able to impose tight bounds on the non-diagonal elements of $U_R$.

\subsection{Constraints on \texorpdfstring{$U_R$}{UR} from \texorpdfstring{$Z$}{Z} boson decays}\label{Zdecay0}

In terms of the previous analysis, a sizable $Y_{1}^e$ or $Y_{3}^e$ is dangerous, as it causes $U_R$ to deviate significantly from the unit matrix and unlike $U_L$, this leaves strong imprints on the neutral current coupling to $Z$. The reason is attributed to $L_R$, which carries a hypercharge different from that of the SM right-handed leptons and thus gives rise to a large correction to the gauge coupling of the $Z$ boson with the charged leptons, 
\begin{equation}\label{eq:NCofe}
\mathcal{L}_{NC} \supset 
\frac{g}{c_{W}}
Z_{\mu} (g^{Z}_{R})_{ij} \bar{ {e}}_{Ri}\gamma^{\mu}  {e}_{Rj} ~~ {\rm with} ~~\left(g^{Z}_{R}\right)_{ij}=(s_{W}^{2})\delta_{ij}+  (\delta g^{Z}_{R})_{ij},
\end{equation}
where $c_{W}=\cos \theta_{W}$ and $s_{W}=\sin \theta_{W}$ with $\theta_{W}$ the Weinberg angle. In the above couplings, the first term denotes the prediction of the SM; the deviation (second term) is due to the mixing between the SM right-handed charged leptons and $L_R$,
\begin{align}
  \delta (g^{Z}_{R})_{ij} &= 
  (U^{\dagger}_R)_{i4}(-1/2)(U_R)_{4j} \,.\label{eq:zemu}
\end{align}
The deviation is the second order of the mixing elements $(U_R)_{4i}$, of which the largest is $(U_R)_{41}$ in our setup. Hence, the most significant deviations that only involve the light leptons are $ \delta (g^{Z}_{R})_{2 2}$ and $\delta (g^{Z}_{R})_{2 1} $, which are highly constrained by the precise data on $Z$ boson decays.

In general, the branching ratios of $Z$ boson decaying into a pair of charged leptons are written as $ \textrm{Br}(Z \to \ell_{i}^{\pm} \ell_{j}^{\mp}) = \frac{1}{2} [\Gamma(Z \to \ell_{i}^{+} \ell_{j}^{-}) + \Gamma(Z \to \ell_{j}^{+} \ell_{i}^{-})]/{\Gamma_Z}$, and the LEP Electroweak Working Group gives the total width of the $Z$ boson $\Gamma_Z = 2.4955 \pm 0.0023 \; {\rm GeV}$~\cite{ParticleDataGroup:2024cfk,ALEPH:2005ab}. The decay width at tree level is expressed as
\begin{equation}\label{Zdecay}
\begin{split}
    \Gamma(Z\to \ell_{i}^+ \ell_{j}^-) =  \frac{G_{F} M_Z^3}{3 \sqrt{2} \pi} [(g_L^Z)_{ij}^2 + (g_R^Z)_{ij}^2]
    \left(1-\frac{m^2}{M_Z^2}\right)^2 \left(1+\frac{m^2}{2M_Z^2}\right), 
\end{split}
\end{equation}
with $m = \max\{m_i, m_j\}$, which can be safely neglected. In calculating, we take ${G}_{F}\approx 1.18\times 10^{-5}\;{\rm GeV}^{-2}$ and $M_Z\approx91.19\;{\rm GeV}$. 
In our model, $(g_L^Z)_{ij}$ take their SM values and do not receive corrections, $(g_L^Z)_{ij} = (-1/2+s_{W}^{2})\delta_{ij}$, while $g_R^Z $ receive modifications as Eq.~\eqref{eq:zemu}.

\subsubsection{Constraints on \texorpdfstring{$\delta (g^Z_R)_{ii}$}{gRii} from Lepton Flavor Universality (LFU) tests}

The $i'$th flavor of the charged lepton mixes with $L_R^-$ and thus modifies the flavor-conserving couplings $(g_R^Z)_{ii}$ by $|U_{4i}|^2$ (see Eq.~\eqref{eq:zemu}), which is not universal to various flavors and then leads to LFU violation of $Z$ boson decay. Concretely, for the $Z$ decay into some flavor $\ell_i$, the deviation from the SM prediction is expressed as the ratio
\begin{align}\label{eq:leptondecaywidthofZ}
\f{\Gamma(Z \to \bar\ell_{i} \ell_{i})}{ \Gamma(Z \to \bar\ell_{i} \ell_{i})_{\rm SM}} =  \frac{(-1/2+s_{W}^{2})^2+\left(s_{W}^{2}-\frac{1}{2}(U_{R})^*_{4 i}(U_R)_{4 i}\right)^2}{(-1/2+s_{W}^{2})^2+(s_{W}^{2})^2}
\approx  1-1.8283|(U_R)_{4i}|^2 \,.
\end{align}
The current measurements of LFU of $Z$ boson decay are presented as the following~\cite{ALEPH:2005ab}
\begin{equation}
    \begin{split}  \label{eq:LFUofZ_tau&mu}
    \f{\textrm{Br}(Z\to \mu^+ \mu^-)}{\textrm{Br}(Z\to e^+ e^-)} = 1.0009 \pm 0.0028 \,,\quad 
    \f{\textrm{Br}(Z\to \tau^+\tau^-)}{\textrm{Br}(Z\to e^+ e^-)} = 1.0019 \pm 0.0032 \,,
    \end{split}
\end{equation} 
with a correlation of $+0.63$.

Consider the two mixing patterns that will be studied to solve the muon $g-2$ anomaly, corresponding to $Y_1^e\neq0$ and $Y_3^e\neq0$ in Eq.~\eqref{Me:full}, respectively. For the former case, both the branch ratios of the $e^+e^-$ and $\mu^+\mu^-$ channels may be sizably decreased, and then we have the following approximation to LFU
\begin{align}
\begin{split}
    \f{\textrm{Br}(Z\to \mu^+ \mu^-)}{\textrm{Br}(Z\to e^+ e^-)} &\approx  1-1.8283(|(U_R)_{4 2}|^2-(U_R)_{4 1}|)^2
    \,, \\
    \f{\textrm{Br}(Z\to \tau^+\tau^-)}{\textrm{Br}(Z\to e^+ e^-)} &\approx 1+1.8283|(U_R)_{4 1}|^2. 
\end{split}
\end{align}
The scenarios $Y_1^e\neq 0$ and $Y_3^e\neq 0$, respectively, give a hierarchy between the two elements $|(U_R)_{41}|, |(U_R)_{43}|\gg |(U_R)_{42}|$. Therefore, applying the data in Eq.~\eqref{eq:LFUofZ_tau&mu}, we get the constraint $|(U_R)_{4 1}| \leq 0.053$. 
For the latter case, instead the branch ratios of the $\mu^+\mu^-$ and $\tau^+\tau^-$ may be modified and then one has
\begin{equation}
\begin{split}
    \f{\textrm{Br}(Z\to \mu^+ \mu^-)}{\textrm{Br}(Z\to e^+ e^-)} \approx  1-1.828|(U_R)_{4 2}|^2     \,,  \quad 
    \f{\textrm{Br}(Z\to \tau^+\tau^-)}{\textrm{Br}(Z\to e^+ e^-)}\approx 1-1.828|(U_R)_{4 3}|^2  \,.
\end{split}
\end{equation}
Then, Eq.~\eqref{eq:LFUofZ_tau&mu} imposes the bounds on the individual mixing elements, $|(U_R)_{4 2}| \leq 0.032 $, and $|(U_R)_{4 3}| \leq 0.027 $.

\subsubsection{Constraints on \texorpdfstring{$\delta (g^Z_R)_{i\neq j}$}{gRij} from cLFV tests}

The cLFV decay of the $Z$ boson imposes a strong bound on the non-diagonal couplings in Eq.~\eqref{eq:zemu}. The upper bounds of cLFV decays of the $Z$ boson are listed as~\cite{ATLAS:2014vur,ATLAS:2021bdj}
\begin{equation}
\begin{split}\label{eq:zij}
    \textrm{Br}(Z \to e^{\pm} \mu^{\mp}) &\leq 7.5 \pm 10^{-7} \,,\\
    \textrm{Br}(Z \to e^{\pm} \tau^{\mp}) &\leq 5 \pm 10^{-6} \,,\\
    \textrm{Br}(Z \to \mu^{\mp} \tau^{\mp}) &\leq 6.5 \pm 10^{-6} \,. 
\end{split}
\end{equation}
The first and third bounds are relevant to our discussions. The first imposes a constraint to the $Y_1^e\neq0$ scenario, where the $Z\to e^\pm\mu^\mp$ mode has the largest cLFV decay width,     
\begin{equation}
\begin{split}
    \Gamma(Z\to \mu^+ e^-) 
    \approx&\frac{G_{F} M_Z^3}{3 \sqrt{2} \pi} \frac{1}{4} |(U_{R})_{4 1}^* (U_R)_{4 2}|^2 \,,
\end{split}
\end{equation}
obtained from  Eq.\eqref{Zdecay}.  Similarly, the $Z\to \mu^\mp \tau^\pm$ mode restricts the $Y_3^e\neq 0$ scenario.

Note that these cLFV decays arise at the quartic order of mixing, and consequently their constraints on the individual mixing may be weaker than those from the LFU tests, which are sensitive to the quadratic order of mixing. Concretely, cLFV of Eq~\eqref{eq:zij} gives  
\begin{align}
    |(U_{R})_{4 1}^* (U_R)_{4 2}| \leq 2.79\times10^{-6}\,, \quad |(U_{R})_{4 3}^* (U_R)_{4 2}| \leq  2.42\times10^{-5}\,.
\end{align}
Due to the hierarchy between the mixing elements mentioned previously, the above bound on the larger element is indeed weaker than the corresponding LFU bound if the hierarchy is sufficiently strong.

\section{Can our LFV flavon resolve the Muon $g-2$ puzzle?}

We have shown that LFV is not present in the SM loops and hence the muon $g-2$ cannot be enhanced by them. However, the flavon, introduced to accommodate full neutrino mixings via the Yukawa term $\ld_2^\ell \bar\ell_2  L_R{\cal F}_\ell^* + h.c.$, retains LFV. This is because, unlike the SM Higgs boson, the flavon merely accounts for the mixing term in the lepton mass matrix. At the same time, this term is potential to lift the value of muon $g-2$, provided that the flavon $s$ is sufficiently light and moreover obtains sizable LFV coupling to either $\mu$ and $e$ or $\mu$ and $\tau$. Note that the latter requirement explains why we turn on $Y_{1}^e$ or $Y_3^e$ in \eqref{Me:full} rather than $M_1^\ell$ or $M_3^\ell$: they generate sizable mixing between $\ell_{1,3 R}$ and $L_R$, which is exactly what we need to produce the desired LFV flavon. In this section, we will try to find out the feasible scenarios of LFV flavon from our flavon-VLL system to resolve the $g-2$ puzzle, confronting many strong constraints related to the precise lepton flavor physics.

\subsection{The profile of the light LFV flavon for muon $g-2$}\label{flavon:emu}

Above all, let us derive the interactions between the flavon and the light charged leptons, from the single term given at the beginning of this section,
\begin{equation}\label{flavon:LFV}
 {\cal L}_{\rm flavon} = -Y^{s}_{ij}\bar{e}_{i} {P}_{R}e_{j} (s+ia)+h.c. ~~{\rm with}~~ Y^{s}_{i j} = \f{\lambda^\ell_2 }{\sqrt{2}} (U_L^*)_{2 i} (U_R)_{4 j},
\end{equation}
with the CP-odd flavon $a$ irrelevant here. From the analytical expression of~\secref{sec:mixing information}, 
one may further approximately express the flavon Yukawa couplings in terms of Lagrangian parameters. We list some of them, which are well approximated as follows
\begin{equation}
\begin{split}\label{eq:analyticys-24}
  Y^s_{22}&=\f{\lambda^\ell_2}{\sqrt{2}}(U_{L}^*)_{22}(U_R)_{42}\approx -\f{\lambda^\ell_2}{\sqrt{2}}\f{a^*|y_\mu|}{(1+|a|^2)^{3/2}},
  \\
  Y^s_{44}&=\f{\lambda^\ell_2}{\sqrt{2}}(U_{L}^*)_{24}(U_R)_{44}
  \approx \f{\lambda^\ell_2}{\sqrt{2}}\f{a^*}{(1+|a|^2)^{1/2}},
  \\
 Y^s_{42}&=\f{\lambda^\ell_2}{\sqrt{2}}(U_L^*)_{24}(U_R)_{42} \approx \f{\lambda^\ell_2}{\sqrt{2}}\f{a^*|a y_\mu|}{(1+|a|^2)^{3/2}},
 \\
 Y^s_{24}&=\f{\lambda^\ell_2}{\sqrt{2}}(U_L^*)_{22}(U_R)_{44} \approx -\f{\lambda^\ell_2}{\sqrt{2}}\f{a^*/|a|}{(1+|a|^2)^{1/2}}. 
\end{split}
\end{equation}%
They are for the case where the $b=Y_{1,3}^e/m_L$ parameter is negligible. According to our ansatze, $Y^{s}_{2 4} \simeq \f{1}{\sqrt{2}}\lambda^\ell_2\sim{\cal O}(1)$, and thus it is of interest to study whether the $s-e_4-\mu$ loop could fill the muon $g-2$ discrepancy, but later analysis negates this possibility. Moreover, diagonal coupling $s\mu^+\mu^-$ is heavily suppressed.

Therefore, to enhance the muon $g-2$ value, we will rely on the sizable LFV coupling in two scenarios: I) $Y^{s}_{21}$ and/or $Y^{s}_{12}$; II) $Y^{s}_{23}$ and/or $Y^{s}_{32}$. They correspond to $Y^e_{1}\neq0$ and $Y^e_{3}\neq0$  in Eq.~(\ref{Me:full}), respectively. In some parameter space, which is reliable in our discussion, an approximate approach can be applied to diagonalize the $3\times 3$ charged mass matrix (with the full details cast in Appendix~\secref{sec:mixing information}), to get the expression for the $Y_1^e\neq0$ case,
\begin{equation}
\begin{split}\label{eq:analyticys-21}
Y^s_{21}& = \f{\lambda^\ell_2}{\sqrt{2}}(U_{L}^*)_{22}(U_R)_{41} \approx \f{\lambda^\ell_2}{\sqrt{2}}  \f{a^*/|a|}{\sqrt{1 + |a|^2}}\f{b |y_\mu|}{\sqrt{|y_\mu|^2 + |a b|^2}} \,,  \\
Y^s_{12}& = \f{\lambda^\ell_2}{\sqrt{2}}(U_{L}^*)_{21}(U_R)_{42} \approx \f{\lambda^\ell_2}{\sqrt{2}} \f{a^* y_e |b|}{(|y_\mu|^2 + |a b|^2)^{3/2}} \f{|a|(|y_\mu|^2-|b|^2)}{1 + |a|^2} \,.
\end{split}
\end{equation}
The former is on the order $\lambda_2^\ell|b|$, while the latter is further suppressed by the Yukawa coupling of the electron $y_e\equiv Y_{11}^ev_h/m_L\simeq m_e/m_L\sim {\cal }{O}(10^{-6})$, thus playing a minor role.

In the $s\bar\mu e$ scenario, we will see that a very light flavon can enhance muon $g-2$, but the enhancement is almost saturated at $m_s\sim m_\mu$. Therefore, $m_s$ of interest is set to lie above $m_\mu$. In fact, an even lighter flavon will open the decay channel $\mu\to e+ s$ at tree level, which is strongly constrained by precise muon measurements such as its lifetime. Therefore, our discussion is restricted to the case $m_s>m_\mu+m_e$, and then the decay of the flavon is dominated by the $e\mu$ mode~\footnote{A fine-tuned flavon mass within the very narrow window $m_\mu-m_e<m_s<m_\mu+m_e$ is considered in~\cite{Ema:2022afm}; in this case, $\phi$ is long-lived given that it has no other decay channels.}. We have a similar conclusion for the flavon in the $s\bar\mu\tau$ scenario. 

The LFV flavon also has a small diagonal coupling with the electron, and this  induced coupling is estimated to be
\begin{equation}\label{eq:analyticys-11}
Y^s_{1 1} = \f{\lambda^\ell_2}{\sqrt{2}} (U_{L}^*)_{21} (U_R)_{41} \approx -\f{\lambda^\ell_2}{\sqrt{2}}  \f{y_e a^* |y_\mu| |b|^2}{(|y_\mu|^2 + |a b|^2)^{3/2}} \,.
\end{equation}
Then, we should concern ourselves whether that light flavon is excluded by electron beam dump experiments that produce the light flavon via a Bremsstrahlung-like process. However, note that the experimentally searched events are produced by long-lived scalars flying over macroscopic
distances and decaying back to electron pairs, thus imposing no constraints on the flavon in our model which decays into $e+\mu$ instantly. By the way, the astrophysical constraints for the light bosons coupling to an electron are also irrelevant to the LFV flavon under consideration, since its mass is heavier than $m_\mu$~\cite{Hardy:2016kme}. Anyway, in principle, $Y_{11}^s$ can be made as small as will, as long as the full neutrino mixings, which require a properly large $|a|$, are achievable. However, according to a quite recent study published in arXiv~\cite{Li:2025beu}, the supernova may impose a  bound on the $m_\mu$ scale LFV flavon with properly coupling to $e\mu s$. 

\subsection{Muon \texorpdfstring{$g-2$}{g-2} from the LFV flavon loop}

\begin{figure}[ht]  
\centering  
    {
    \begin{tikzpicture}[line width=1.0 pt, scale=0.6, >=latex]
    \begin{feynhand}
    \vertex [label=left:$\ell_{i}$] (a) at (-6,0) ;
    \vertex [label = below:$Y^{s}_{f i} P_R + (Y^{s})^*_{i f} P_L$] (b) at (-3,0) ;
    \vertex (c) at (0,0) ;
    \vertex [label = below:$Y^{s}_{j f} P_R + (Y^{s})^*_{f j} P_L$] (d) at (3,0) ;
    \vertex [label=right:$\ell_{j}$] (e) at (6,0) ;
    \vertex [label=right:$\gamma$] (f) at (0,-4) ;
    \propag  [fermion] (a) to (b);
    \propag  [fermion] (b) to (c);
    \propag  [fermion] (c) to (d);
    \propag  [fermion] (d) to (e);
    \draw [sca] (3,0) arc (0:180:3);
    \propag  [boson] (c) to (f);
    \node at (0,3.4) {$s$};
    \node at (-1,0.5) {$\ell_{f}$};
    \end{feynhand}
    \end{tikzpicture}
    }
    \captionsetup{justification=centering}
    \caption{The LFV flavon loop induced charged lepton $g-2$ for $\ell_i = \ell_j $ and charged lepton flavor violating decay $\ell_i \to \ell_j \gamma$ for $\ell_i \not= \ell_j$ with intermediate leptons $\ell_f$.
    }
    \label{fig:feyLFVsitoj}
\end{figure}
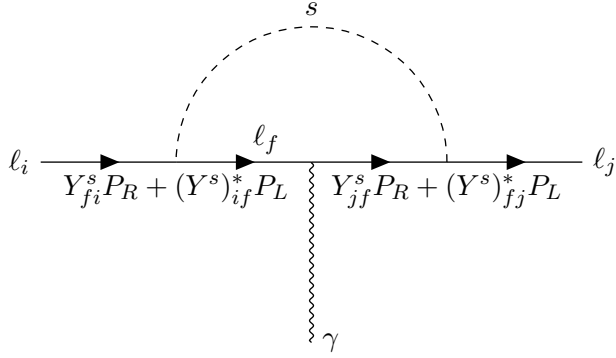
In this subsection, we will investigate the muon $g-2$ via the flavon loops, first giving the no-go via the $s-\mu-e_4$ loop in the minimal case with only $a=M_2^\ell/m_L\neq 0$. Then we analyze how the LFV loop $s-e-\mu$, which benefits in free of any large mass scale suppression, may work at the price of one extra mixing parameter $Y_1^e$. Prior to those concrete analyses, we will have a general discussion on the formulas of $g-2$ and further analyze the behaviors of the loop function in two limits, the decoupling limit with one large mass scale (compared to the mass scales of the external legs) in the loop and the opposite case with only light loop mass scale. These discussions are also useful in  cLFV decay, which shares the same quantum process as the charged lepton $g-2$.  

\subsubsection{Preparation: loop functions}

To keep the discussion most general, let us start from the effective Lagrangian of a LFV CP-even scalar boson coupling to different SM charged leptons $\ell_{i}$, which has been given in Eq.~\eqref{flavon:LFV}. Then, following the calculation in Ref.~\cite{Lynch:2001zr,Lynch:2001zs}, the one loop contribution to the $g-2$ of $\ell_{i}=\{e,\,\mu,\,\tau\}$ via the LFV scalar boson loop $\ell_i-\ell_f-\ell_i$ (such as \figref{fig:feyLFVsitoj}) is calculated to be
\begin{equation}
\begin{split}\label{eq:}
    \delta a_{\ell_i}^{s}(\ell_f) = \f{m_{\ell}^2}{16 \pi^2 m_s^2} \left[(|Y^{s}_{i f}|^2 + |Y^{s}_{f i}|^2) 
    F_s \left(\frac{m_f^2}{m_{s}^2}, \frac{m_{i}^2}{m_s^2}\right)
    +
     \f{m_f}{m_{i}} {\rm Re}(Y^{s}_{i f}Y^{s}_{f i}) G_s \left(\frac{m_f^2}{m_{s}^2}, \frac{m_{i}^2}{m_s^2}\right)\right] \,.
\end{split}
\end{equation}
In the above expression, the first part does not require chirality flipping of the loop fermion, while the second part does, and their loop functions are respectively given by
\begin{equation}
\begin{split}\label{eq:lfint}
F_s(x, y) = & \int_{0}^{1} \mathrm{d}z \f{z (1 - z)^2}{z + (1-z) x - z(1 - z) y} \,, \\
G_s(x, y) = & \int_{0}^{1} \mathrm{d}z \f{2 (1 - z)^2}{z + (1-z) x - z(1 - z) y} \,.
\end{split}
\end{equation}
Their variables are $x\equiv {m_f^2}/{m_{s}^2}$, $y\equiv{m_{i}^2}/{m_s^2}$, the square of the ratio of the internal and external fermion mass to the flavon mass, respectively; in our work, we always have $y>1$, namely the loop never develops an imaginary part.

\begin{figure}[ht]
    \centering
    {
    \includegraphics[width=0.29\linewidth]{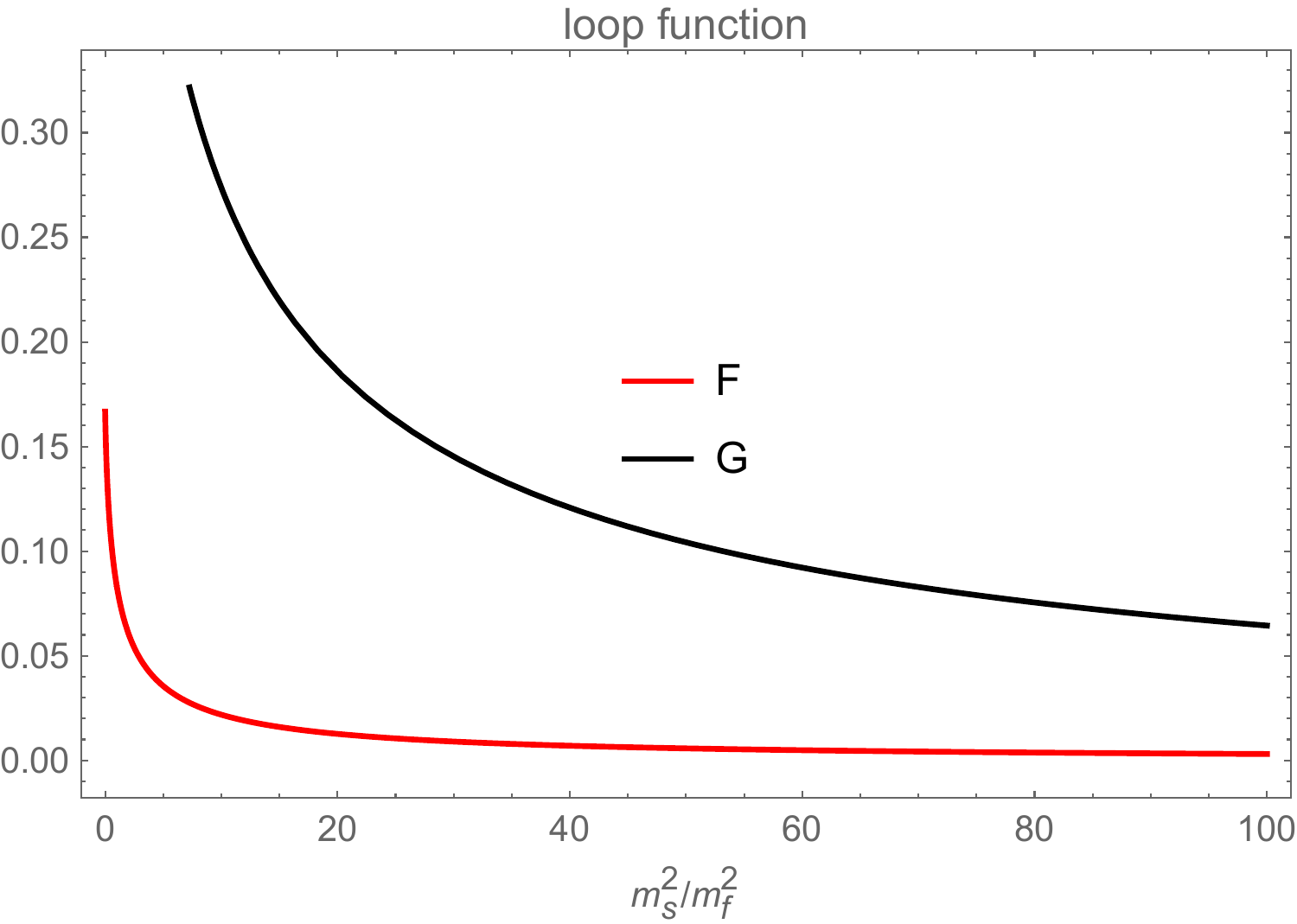}
    }
    {
    \includegraphics[width=0.32\linewidth]{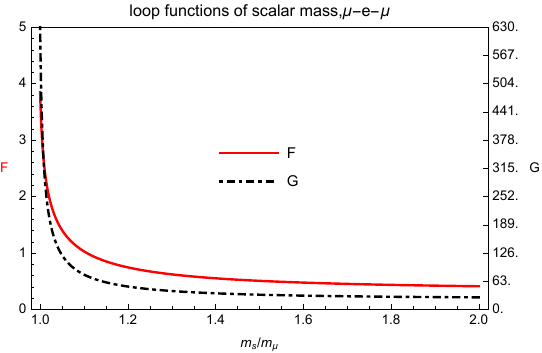}
    }
    {
    \includegraphics[width=0.32\linewidth]{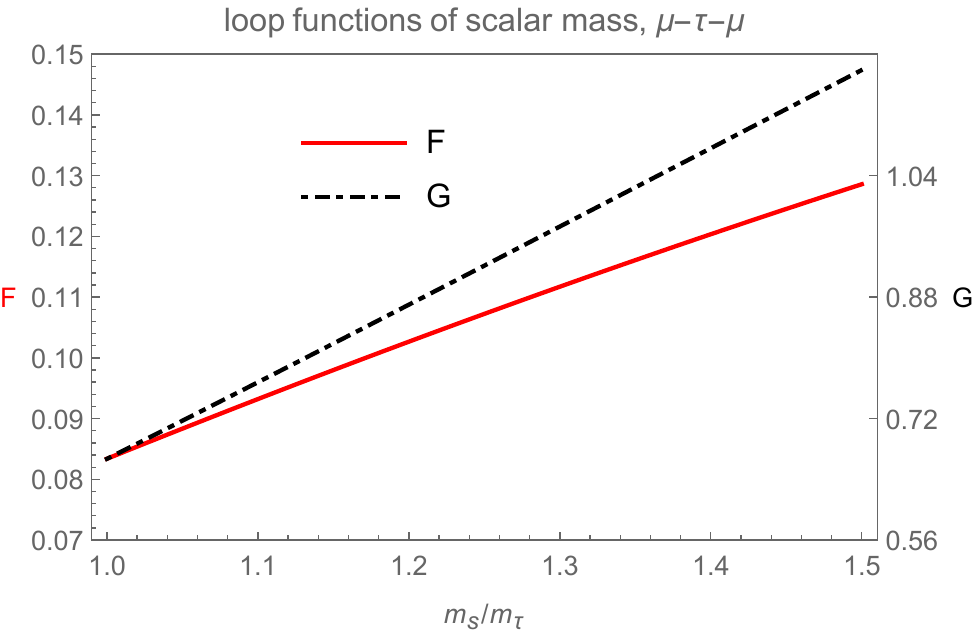}
    }
    \captionsetup{justification=raggedright, singlelinecheck=false} 
    \caption{Loop functions associated with the charged radiative process with the CP-even flavon running in the loop. The left panel is for $F_s \left({m_{s}^2}/{m_f^2}\right)$ and $G_s \left({m_{s}^2}/{m_f^2}\right)$ given in Eq.\eqref{eq:lfs}. For later consideration, we plot the loop functions for the $s\bar \mu e$ and  $s\bar \mu \tau$ scenarios in the middle and right panels, respectively. }
    \label{fig:lf2ver}
\end{figure}

In general, the loop functions $F_s(x,y)$ and $G_s(x,y)$ do not have simple analytical expressions. To demonstrate the behaviors of these loop functions, on the left of~\figref{fig:lf2ver}, we show them in a special limit (for instance, it is the case in the $s\bar\mu\tau$ scenario studied in section.~\ref{smutau}) where the external fermion mass is negligible compared with the flavon mass $m_s$, giving $y\to 0$. In this limit, the denominator of the integral can be reduced to be $z + (1-z) x$ both for $F_s$ and $G_s$, and then the integral can be implemented explicitly to get the widely used expression.
\begin{align}\label{eq:lfs}
F_{s}(x)&=-\frac{x^{3}-6x^{2}+3x+6x\log x+2}{6(1-x)^{4}},\quad
G_{s}(x) =\frac{x^{2}-4x+2\log x+3}{(1-x)^{3}}.
\end{align}
We always have $|F_s/G_s| < 1$, which is explicit in the first panel of~\figref{fig:lf2ver}. But if the scalar boson becomes as light as the external fermion, i.e., $y \to 1$, then we do not have a simple expression like the above, and one should use the complete numerical integral. We will encounter this limit in the $s\bar\mu e$ flavon scenario. For illustration, we give the numerical sample for a special case: $F_s \left(\frac{m_e^2}{m_{\mu}^2}, \frac{m_{\mu}^2}{m_{\mu}^2}\right) = 3.85022   \,,~~
G_s \left(\frac{m_e^2}{m_{\mu}^2}, \frac{m_{\mu}^2}{m_{\mu}^2}\right) = 629.236$. 

\subsubsection{The sharpest razor: insufficient enhancement from the \texorpdfstring{$\mu-e_4-s$}{mu-e4-s} loop}

It is tempting to consider whether the minimal case with $Y^{e}_{1, 2, 3} = 0$ and $M^{\ell}_{1, 3} = 0$ in \eqref{Me:full} can provide a viable solution to the muon $g-2$ puzzle. We found that the flavon-induced heavy mediator contribution is indeed capable of doing that job as long as $e_4$ is as light as the weak scale. In the following, a simple analytical analysis could lead us to this conclusion.

In this case, the contribution of flavon to $(g-2)_{\mu}$ is dominated by the $\mu$-$e_4$-$s$ loop. For $m_\mu\ll m_s<m_L$ one can use Eq.~\eqref{eq:lfs}, and then $G_{s}(x)\approx-\frac{1}{x}$ on the order of $10^{-3}$; moreover,  $1/8<F_s(x)/G_s(x)<1/6 $.  Therefore, the contribution can be approximated to be
\begin{align}
\delta a_{\mu}^{s}(e_4)
=& \frac{\kappa_{m}^2}{16\pi^{2}} \left(\frac{1}{6}(|Y^{s}_{2 4}|^{2}+|Y^{s}_{4 2}|^{2})+\frac{{\rm Re}(Y^{s}_{2 4} Y^{s}_{4 2})}{\kappa_{m}}\right).
\end{align}
where $\kappa_{m}=m_{\mu}/m_{4}\approx |y_\mu|/(1+|a|^2)\ll 1$. It is clearly seen that the relatively light flavon mass becomes irrelevant under this approximation.  Inserting the approximate analytical expression of the chiral rotation and fermion masses, the above result can be expressed in terms of the Lagrangian parameters as
\begin{align}
\delta a_{\mu}^{s}(e_4)
\approx&\frac{\kappa_{m}^2|\lambda^\ell_2|^2}{32\pi^{2}}
\left[\frac{1}{6}\f{(1+|a|^2)^2+|a^2 y_\mu|^2}{(1+|a|^2)^3} - \frac{1}{\kappa_{m}}\f{|a|^2|y_\mu|}{(1+|a|^2)^{2}}\right]\\ 
\nonumber
\approx&\frac{\kappa_{m}^2|\lambda^\ell_2|^2}{32\pi^{2}}
\f{1/6-|a|^2}{1+|a|^2}
\approx 3.16629\times10^{-9}\f{1/6-|a|^2}{1+|a|^2}.
\end{align}
where the $y_\mu$ term is neglected compared with $a$ in the final expression. It is a good approximation under numerical tests. We have taken $|\lambda^\ell_2|=1$ and $\kappa_{m} \simeq 10^{-3}$, namely a weak-scale heavy lepton, to make the final numerical estimation. Therefore, whether this solution survives strongly depends on whether that light $e_4$ has been excluded. Because $\lambda_2^\ell$ is large, the dominant decay channel of $e_4$ is $\mu+s$ with $s\to \mu^+\mu^-$, and as a result the light $e_4$ has definitely been ruled out by the $6\mu$ signal at LHC.

\subsubsection{The first sight: \texorpdfstring{$Y^{e}_{1}\not=0$}{Ye1not=0} and the \texorpdfstring{$\mu-e-s$}{mu-e-s} loop enhanced by a light flavon}

Contrary to the chiral flip via the large mass of the heavy lepton $e_4$, the contribution of the loop such as $\mu-e-s$, $\mu-\tau-s$ and $\mu-\mu-s$ can be enhanced through eliminating the large mass scale suppression in the loop, given that the flavon mass lies much below the weak scale. Here, the flavon is supposed to dominantly couple to leptons violating the lepton flavors by virtue of the symmetry selection rule, and consequently, the $\mu-\mu-s$ is supposed to only play a negligible role.

To realize such a scenario in our model, we have to turn on one more parameter, $Y^{e}_{1}\not=0$ or $Y^{e}_{3}\not=0$, and then either $(U_R)_{41}$ or $(U_R)_{43}$ gains a sizable value, as discussed in Appendix~\secref{sec:mixing information}. The $\mu-e-s$ loop is more promising than the $\mu-\tau-s$ loop since $m_\tau\gg m_\mu$, which may incur suppression again, and thus let us focus on the former loop first, whose contribution is given by
\begin{equation}\label{eq:aes}
\begin{split}
  \delta a_{\mu}^{s}(e) 
    = &
    \f{m_{\mu}^2}{16 \pi^2 m_s^2} \left[(|Y^{s}_{2 1}|^2 + |Y^{s}_{1 2}|^2) F_s \left(\frac{m_e^2}{m_{s}^2}, \frac{m_{\mu}^2}{m_{s}^2}\right) + \f{m_e}{m_{\mu}} {\rm Re}(Y^{s}_{2 1} Y^{s}_{1 2})G_s \left(\frac{m_e^2}{m_{s}^2}, \frac{m_{\mu}^2}{m_{s}^2}\right)\right]\,.
    \end{split}
\end{equation}
For the sake of maximal enhancement, we consider the flavon mass around the order of the muon mass. Then, one should  use the complete loop integral in Eq.~\eqref{eq:lfint} and the corresponding behaviors of $F_s$ and $G_s$ are shown in the center panel of~\figref{fig:lf2ver}. 

Seemingly, there is no suppression in the loop functions from the heavy fermion scale, but it is hidden in the effective Yukawa couplings by means of heavy-light mixings. Therefore, we need to carefully analyze this suppression to see if it can be compensated by the light flavon enhancement. To that end, we express $\delta a_{\mu}^{s}(e)$ in terms of the Lagrangian parameters. Considering the analytical approximation of Eq.~\eqref{eq:analyticys-21} which shows the hierarchy $\left|Y^{s}_{2 1}\right| \gg \left|Y^{s}_{1 2}\right|$ and thus only the $|Y^{s}_{2 1}|^2$ term is left in Eq.~(\ref{eq:aes}),  
and then the final expression is simplified to be
\begin{equation}
\begin{split}
    \delta a_{\mu}^{s}(e) \approx 
    |\f{b}{y_\mu}|^2\f{|\ld_2^\ell|^2}{32 \pi^2} \f{m_{\mu}^4}{m_L^2 m_s^2} F_s \left(\frac{m_e^2}{m_{s}^2}, \frac{m_{\mu}^2}{m_{s}^2}\right)\,.
    \end{split}
\end{equation}
The suppression of heavy-fermion scales as $m_\mu^2/m_L^2\sim 10^{-8}$ for a TeV scale $m_L$; taking $m_s=m_\mu$ for a numerical estimation which leads to $F_s\simeq 3.8$, then the required size can be easily obtained as long as $b$ is a few times larger than $y_\mu$. For comparison, the resulting $\mu-\mu-s$ loop contribution can be estimated as
\begin{equation}\label{eq:amus}
\begin{split}
  \delta a_{\mu}^{s}(\mu) \approx
    \frac{|\ld_2^\ell|^2}{32\pi^{2}} \frac{m_\mu^4}{m_s^2 m_L^2} \f{|a|^2}{(1+|a|^2)^2}\left[ 2F_s \left(\frac{m_{\mu}^2}{m_{s}^2}, \frac{m_{\mu}^2}{m_{s}^2}\right) + G_s \left(\frac{m_{\mu}^2}{m_{s}^2}, \frac{m_{\mu}^2}{m_{s}^2}\right)\right]\,.
\end{split}
\end{equation}
As expected, even if $|a|$ is as large as the order one, it is still much smaller than $\delta a_\mu^s(e)$. This is attributed to the mixing pattern that gives $|(U_R)_{4 1}|$ considerably larger than $|(U_R)_{4 2}|$. 

Our solution to the muon $g-2$ puzzle turns to a light flavon and its LFV couplings, which inevitably lead to the cLFV process with a rate strongly related to the size of the muon $g-2$. Within the SM, cLFV decays have extremely small branching ratios ($\ll 10^{-50}$) even when accounting for the measured neutrino mass differences and mixing angles. Therefore, such types of rare decays provide ideal laboratories for searching for new physics giving rise to cLFV, and the null results impose stringent constraints on the related new physics. Therefore, one challenge to our scenario is whether it is tolerated by the cLFV constraints, and we will discuss them in the following.

\subsection{The LFV flavon resolution against the cLFV constraints }\label{sec:LFV processes}

We carry out the discussion on cLFV of $\mu$ to $e$ conversion as an example, and most results can be used in the case $\tau$ to $\mu$ after simple replacement in the corresponding expressions.

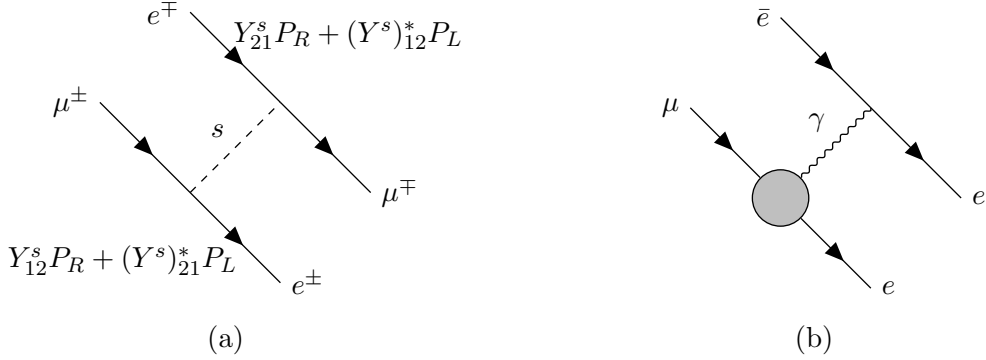
\begin{figure}[ht]  
    \centering  
    \subfloat[]
    {
    \begin{minipage}{0.46\linewidth}
        \centering
    \begin{tikzpicture}[line width=1.0 pt, scale=0.6, >=latex]
    \begin{feynhand}
    \vertex [label=left:$\mu^{\pm}$](a) at (-3,2) ;
    \vertex [label=right:$e^{\pm}$](b) at (1,-2) ;
    \vertex [label=right:$\mu^{\mp}$](c) at (3,0) ;
    \vertex [label=left:${e}^{\mp}$](d) at (-1,4) ;
    \vertex (e) at (1,2) ; 
    \vertex (f) at (-1,0) ;
    \propag  [fermion] (a) to (f);
    \propag  [fermion] (f) to (b);
    \propag  [fermion] (d) to (e);
    \propag  [fermion] (e) to (c);
    \propag  [sca] (f) to [edge label = $s$](e);
    \node at (-2.5,-1.5) {$Y^{s}_{1 2} P_R + (Y^{s})^*_{2 1} P_L$};
    \node at (2.5,3.5) {$Y^{s}_{2 1} P_R + (Y^{s})^*_{1 2} P_L$};
    \end{feynhand}
    \end{tikzpicture}
    \end{minipage}
    }
    \subfloat[]
    {
    \begin{minipage}{0.46\linewidth}
        \centering
    \begin{tikzpicture}[line width=1.0 pt, scale=0.6, >=latex]
    \begin{feynhand}
    \vertex [label=left:$\mu$](a) at (-3,2) ;
    \vertex [label=right:$e$](b) at (1,-2) ;
    \vertex [label=right:$e$](c) at (3,0) ;
    \vertex [label=left:$\bar{e}$](d) at (-1,4) ;
    \vertex (e) at (1,2) ; 
    \vertex [grayblob] (f) at (-1,0) {} ;
    \propag  [fermion] (a) to (f);
    \propag  [fermion] (f) to (b);
    \propag  [fermion] (d) to (e);
    \propag  [fermion] (e) to (c);
    \propag  [boson] (f) to [edge label = $\gamma$](e);
    \end{feynhand}
    \end{tikzpicture}
    \end{minipage}
    }
    \captionsetup{justification=raggedright, singlelinecheck=false} 
    \caption{Left: The $\Delta L_{\mu, e}= \pm2$ process of oscillation between muonium and antimuonium. Right: The $\Delta L_{\mu, e}= \pm1$ process of $\mu\to 3e$, dominated by dipole operator; it also receives a tree level contribution such as the left figure with one final state $\mu$ replaced by $e$.}
    \label{fig:feyLFVs}
\end{figure}

Three types of $\mu\to e$ processes are experimentally detected: radiative decay $\mu \to e \gamma$, three-body decay $\mu \to 3 e$, and $\mu-e$ conversion in nuclei $\mu N \to e N$. In our model, $\mu \to 3 e$ presents two contributions as shown in~\figref{fig:feyLFVs}: the tree-level channel mediated by the LFV flavon and the radiative channel mediated by the photon; the former can be neglected here due to the smallness of the flavon coupling to the electron $Y_{11}^s$, and the radiative channel is dominant by photon mediation. Then, all these $\mu\to e$ processes can be described by the following dimension-5 dipole operators,
\begin{align}
\mathcal{L} & = c_{L} \frac{e}{8 \pi^2} m_\mu \left(\bar{e}\, \sigma^{\alpha \beta} P_{L}\, \mu \right) F_{\alpha \beta} + c_{R} \frac{e}{8 \pi^2} m_\mu \left(\bar{e}\, \sigma^{\alpha \beta} P_{R} \, \mu \right) F_{\alpha \beta} + h.c. \, ,
\label{eq:dipolemutoe} 
\end{align}
which are valid at the scale of muon mass. The Wilson coefficients $c_{L, R}$ are model dependent and are calculated by the matching at low energy. Specific to our model, they are calculated to be
\begin{align}
\begin{split}\label{eq:WcLRmu+e} 
c_{L} & = \frac{1}{4 m_{\mu}} \int^{1}_{0} \mathrm{d} x \mathrm{d} y \mathrm{d} z \, \delta(1 - x - y - z)  \\[0.5ex]
&\hspace{8.0em} \left[ \frac{x z m_{e} (Y^{s}_{2 2})^* Y^{s}_{1 2} + y z m_{\mu} Y^{s}_{2 2} (Y^{s}_{2 1})^* + (x + y) m_{\mu} (Y^{s}_{2 2})^* (Y^{s}_{2 1})^*}{z m_{s}^2 - x z m_{e}^2 - y z m_{\mu}^2 + (x + y) m_{\mu}^2 - x y q^2} \right.  \\[0.5ex]   
&\hspace{8.0em} \left. + 
\frac{x z m_{e} (Y^{s}_{2 1})^* Y^{s}_{1 1} + y z m_{\mu} Y^{s}_{1 2} (Y^{s}_{1 1})^* + (x + y) m_{e} (Y^{s}_{2 1})^* (Y^{s}_{1 1})^*}{z m_{s}^2 - x z m_{e}^2 - y z m_{\mu}^2 + (x + y) m_{e}^2 - x y q^2} \right],
    \end{split}
\end{align}
where the two terms in the square bracket respectively denote that the leptons running in the loop are the muon and electron, and the latter is subdominant because $|Y_{11}^s|\ll |Y_{22}^s|$. $c_R$ is obtained from $c_L$ where the flavor indices are swapped in each Yukawa coupling, further taking complex conjugation, e.g., $Y_{12}^s\to (Y_{21}^s)^*$. Likewise for dimension-5 dipole operators related to muon and tau - $c_{L, R}$: $Y_{1 2}^s\to Y_{2 3}^s$, $Y_{2 1}^s\to Y_{3 2}^s$; and $m_\mu \to m_\tau$.

In terms of these operators, now we can calculate the observables. The results are well known, and we quote them in the following. First, the radiative two body decay width of $\mu \to e \gamma$ is
\begin{align}\label{cLFV:1}
\Gamma (\mu \to e \gamma) = \frac{\alpha m_\mu^5}{64 \pi^4} \left(\left|c_R\right|^2 + \left|c_L\right|^2\right) \, . 
\end{align}
Next, at leading in $m_e/m_\mu$, the decay width of three-body cLFV $\mu \to 3 e$ is dominated by the logarithmic term~\cite{Kuno:1999jp,Okada:1999zk}
\begin{equation}
    \Gamma (\mu \to 3 e) \simeq \frac{\alpha^2 m_{\mu}^5}{192 \pi^5} \left|\log\frac{m_{\mu}^2}{m_{e}^2} - 11/4\right| (\left|c_{L}\right|^2 + \left|c_{R}\right|^2) \, .
\end{equation}
Their branch ratios are strongly constrained; we use the PDG data $\Gamma (\mu \to 3 e)/\Gamma < 1.0 \times 10^{-12}$ with $90\%$ C.L. The current upper bounds on cLFV for $\mu$ and $\tau$ are summarized in Table~\ref{tab:my_labelmutau}~\cite{ParticleDataGroup:2024cfk}.
\begin{table}
    \centering
    \begin{tabular}{|c||c|c|}
    \hline
        decay models  & muon &tau \\
        \hline\hline
        mean life & $\tau_{\mu} = 2.1969811 \times 10^{-6} \; {\rm s}$ &  $\tau_{\tau} = 2.903 \times 10^{-13} \; {\rm s}$ \\
        \hline
         cLFV & $\Gamma (\mu \to e \gamma)/\Gamma_{\mu} < 4.2 \times 10^{-13}\;90\%{\rm C.L}$ & $\Gamma (\tau \to \mu \gamma)/\Gamma_{\mu} < 4.4 \times 10^{-8}$ \\
         \hline
         three-body & $\Gamma (\mu \to 3 e)/\Gamma_{\mu} < 1.0 \times 10^{-12}\;90\%{\rm C.L}$ & $\Gamma (\tau \to 3 \mu)/\Gamma_{\tau} < 2.1 \times 10^{-8}$ \\\hline
    \end{tabular}
    \caption{Some PDG data of muon and tau.}
    \label{tab:my_labelmutau}
\end{table}

Finally, the conversion $\mu N \to e N$ is the capture of muon via the Coulomb force of nuclei without the outgoing neutrino, and the observable is defined by the capture rate (CR)
\begin{align}
{\rm CR} (\mu \mathchar`- e, N) = \frac{\Gamma (\mu N \to e N)}{\Gamma_{\rm capt}} \, ,
\end{align}
where $\Gamma (\mu N \to e N)$ is the cLFV $\mu\to e$ conversion rate in some  heavy nuclei $N$, and $\Gamma_{\rm capt}$ is the total capture rate, given in Ref.~\cite{Kitano:2002mt}. 
The conversion rate is calculated to be~\cite{Alonso:2012ji,Abada:2015oba,Abada:2018nio},
\begin{align}
\Gamma (\mu N \to e N) = \left| - \frac{e}{16 \pi^2} c_R D \right|^2 + \left| - \frac{e}{16 \pi^2} c_L D \right|^2 \; ,
\end{align}
where the coefficient $D$ is the form factor whose value can be found in Ref.~\cite{Kitano:2002mt} for various nuclear targets. The strongest bound is imposed by the gold target (concretely, $^{197}_{\;79}{\rm Au}$), which gives $D=0.189$ in units of $m_{\mu}^{5/2}$; the corresponding capture of $\mu$ within SM is through weak interactions, and it is predicted that the rate will be $\Gamma_{\rm capt, \; {\rm Au}} = 13.07 \times 10^6\; {\rm s}^{-1}$~\cite{Suzuki:1987jf}. The experimental upper bound is ${\rm CR} (\mu \mathchar`- e, {\rm Au})<7 \times 10^{-13} \; (90 \% \; {\rm{CL}})$~\cite{SINDRUMII:2006dvw}. In fact, we will find that this bound exceeds the ones from the above two cLFV decays.

\subsection{Kill the \texorpdfstring{$s\bar \mu e$}{s\bar\mue} scenario}

Now, we are in the position to investigate the $s\bar \mu e$ scenario. However, it suffers an additional constraint which should be imposed, the muonium-antimuonium oscillation, which is so strong that the scenario will be closed.

\subsubsection{The muonium-antimuonium oscillation constraint}

The muonium $M$ is a QED bound state of $\mu^+e^-$, and it will oscillate with antimuonium $\overline{M}(\mu^-e^+)$ in the presence of LFV processes with $\Delta L_{e, \mu}=\pm 2$ beyond the SM. Such a phenomenon is one of the goals of the MEG (Mu to Electron Gamma) experiment at the Paul Scherrer Institute (PSI), and the first running provides an upper limit for the oscillation probability $P_{M\overline{M}} \leq 8.2 \times 10^{-11}/S_B$ at 90\% C.L.~\cite{Willmann:1998gd}. This will yield a direct constraint on the LFV coupling $s\bar \mu e$ at tree level shown as in the left of Fig.~\ref{fig:feyLFVs}, without involving other couplings of the flavon as in the cLFV decays discussed previously.

We follow~\cite{Endo:2020mev,Calibbi:2024rcm,Fukuyama:2021iyw} to carry out a detailed study of muonium oscillation in our model where the light flavon cannot be integrated out. As in the $K^0-\overline{K^0}$ and neutrino system, the dynamics of the $M-\overline{M}$ mixing quantum system is described by their mass matrix, which consists of the elements of the effective Hamiltonian $H_{eff}$ among the two states: 
\begin{equation}
     \mathcal{M} =\langle \alpha |H_{eff}|\beta\rangle=
     \begin{pmatrix}
         m-i\Gamma_\mu/2&\Delta m/2\\
         \Delta m/2&m-i\Gamma_\mu/2
     \end{pmatrix}\,,\quad {\rm with} ~~ |\alpha/\beta\rangle=|M~{\rm or}~ \overline{M}\rangle.
\end{equation}
where $m$ and $\Gamma_{\mu}\simeq 3.00\times10^{-19}\;{\rm GeV}$ are the averages of the masses and widths of $M$ and $\overline{M}$, respectively. The off-diagonal element $ \mathcal{M} _{12}=\langle M |H_{eff}|\overline{M}\rangle=\Delta m/2$ is crucial since it induces the $M-\overline{M}$ mass splitting and oscillation. It has been estimated without and with a magnetic field $B$ employed in the experimental apparatus~\cite{Hou:1995np,Hou:1995dg}. 
Then, the time-integrated probability of $M-\overline{M}$ conversion is given by
\begin{equation}\label{eq:muoniumprobablity}
     P_{M\overline{M}} \approx \int^{\infty}_{0}\mathrm{d}t\; \Gamma_\mu \exp(-\Gamma_\mu t)\sin^2(\mathcal{M}_{12} t)=\frac{2\mathcal{M}_{12}^2}{4\mathcal{M}_{12}^2+\Gamma_\mu^2}\,\approx \frac{2\mathcal{M}_{12}^2}{\Gamma_\mu^2}.
\end{equation}
In our model, it is written in terms of the flavon LFV coupling as 
\begin{align}\label{M-barM}
     P_{M\overline{M}} = & \frac{8/(\pi^2 a_B^6 \Gamma_\mu^2)}{(m_{\mu}^2-m_s^2)^2+\Gamma_s^2 m_s^2} 
     \big(\left|c_{0,\,0}\right|^2 \left|(\frac{Y^s_{12}+(Y^s_{21})^*}{2})^2 -(1+\frac{1}{\sqrt{1+X^2}})(\frac{Y^s_{12}-(Y^s_{21})^*}{2})^2\right|^2 \nonumber\\
     &\hspace{0.3em}+
     \left|c_{1,\,0}\right|^2\left|(\frac{Y^s_{12}+(Y^s_{21})^*}{2})^2 -(1-\frac{1}{\sqrt{1+X^2}})(\frac{Y^s_{12}-(Y^s_{21})^*}{2})^2\right|^2\big)\,,
\end{align}
where, as in Ref.~\cite{Calibbi:2024rcm}, we retain the complete propagator of the flavon. In the above formula, $a_B\simeq2.69\times10^{5}\;{\rm GeV}^{-1}$ is the muonium Bohr
radius and $X=6.31\,(B/1{\rm T})$ with $B=0.1\;{\rm T}$ in the MEG, which results in the population probability of the muonium initial state with $(J,m_J)$, $\left|c_{0,\,0}\right|^2=0.32$ and $\left|c_{1,\,0}\right|^2=0.18$~\cite{Hou:1995dg}.   

\subsubsection{The results for the light flavon with \texorpdfstring{$s\bar \mu e$}{sbarmue}}

According to our previous analysis shown in Eq.~(\ref{eq:aes}) and Eq.~(\ref{eq:WcLRmu+e}), the dominant loop that determines the value of muon $g-2$ and the main loop that determines the magnitudes of Wilson coefficients $c_{L,R}$ share similarities both in couplings and loop functions, and consequently, the region favored by the former has the potential to be excluded by the latter. In this subsection, we investigate how a light LFV flavon $s$ in our model can explain the muon deviation $g-2$ when confronted with the cLFV constraint. Although this scenario will be negated by the $M-\overline{M}$ oscillation constraint, the analysis made here will apply to the other scenario that is free of this constraint, and we will present it in another section.

\begin{figure}
    \centering
    {
    \includegraphics[width=0.31\linewidth]{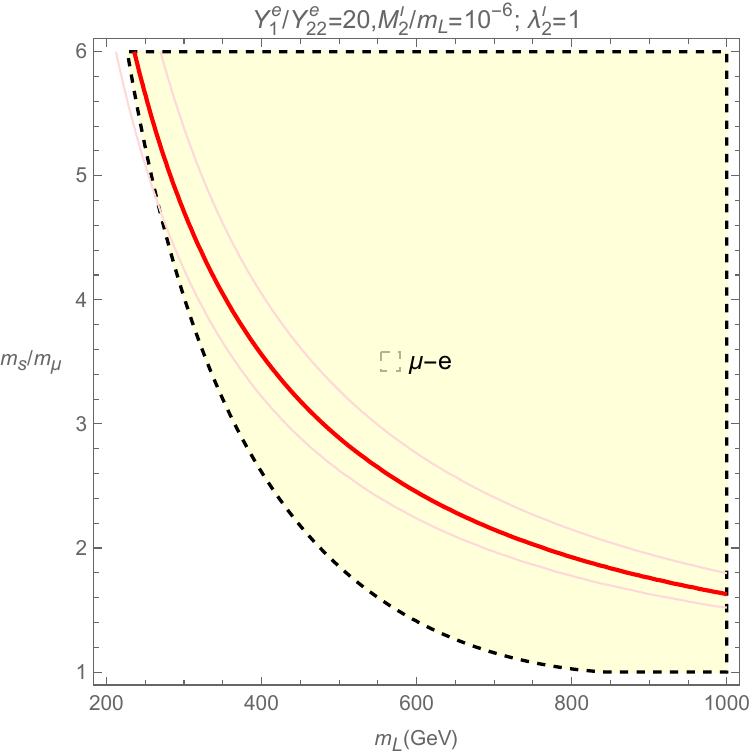}
    }
    {  
    \includegraphics[width=0.31\linewidth]{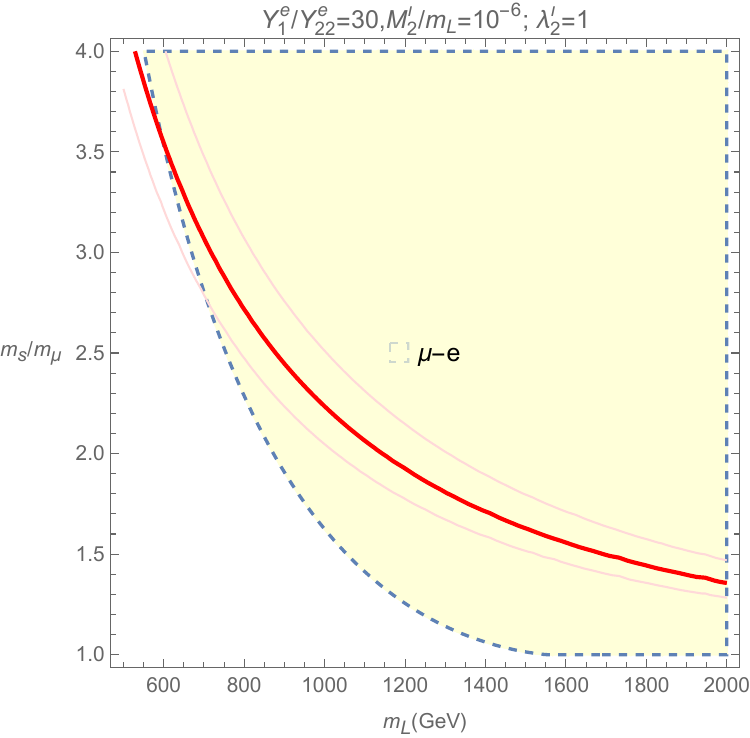}
    }
    {
    \includegraphics[width=0.31\linewidth]{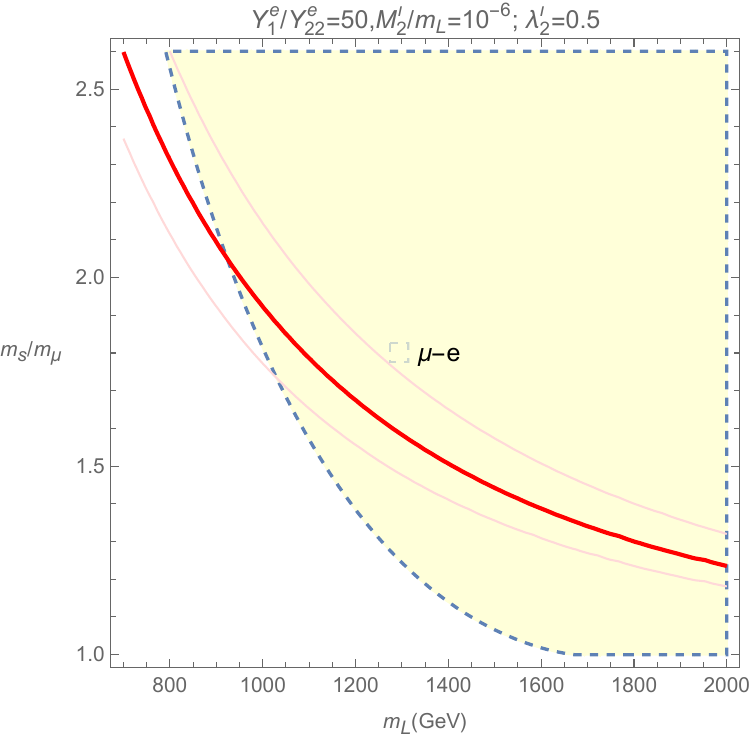}
    }
    \captionsetup{justification=raggedright, singlelinecheck=false} 
    \caption{The light LFV flavon with $e-\mu-s$ loop to explain the muon $g-2$ discrepancy in the $m_L-m_s/m_\mu$ plane (from left to right with $Y_1^e/Y^{e}_{22}=10,30,50$, fixing $M_2^\ell/m_L=10^{-6}$): the 2$\sigma$ confidence band is sandwiched between two thin red lines; only the $e-\mu$ conversion constraint is imposed and allows the yellow shaded region.}
    \label{fig:abab}
\end{figure}  
The first key is to note that $c_{L,R}$ can be additionally suppressed given that $|Y_{22}^s|\ll |Y_{21}^s|$, which leads us to explore the parameter region with a small $a=M^{\ell}_2/m_L\ll 1$. This is because the mixing element $M^\ell_2$ (or $a$) only controls the diagonal Yukawa couplings of the flavon like $Y^s_{22}$; however, the LFV couplings like $Y^s_{2 1, 1 2}$ are proportional to the mixing element $Y^e_{1}$ (or $b$); one can see this explicitly in Eq.~(\ref{eq:analyticys-24}) and Eq.~(\ref{eq:analyticys-21}). The second key is to find that muon $g-2$ benefits from the quasi-degeneracy between the masses of muon and flavon, while $c_{L,R}$ does not. Therefore, the numerical analysis is implemented in the full parameter space characterized as
\begin{equation}
    Y^{e}_{1}/Y^{e}_{22} > 1,\quad  M^{\ell}_{2}/m_L \ll 1, \quad m_s\gtrsim m_{\mu},
\end{equation}
and moreover we limit $m_L\gtrsim 1$ TeV considering collider bound on VLL.

\begin{figure}
    \centering
    {
    \includegraphics[width=0.4\linewidth]{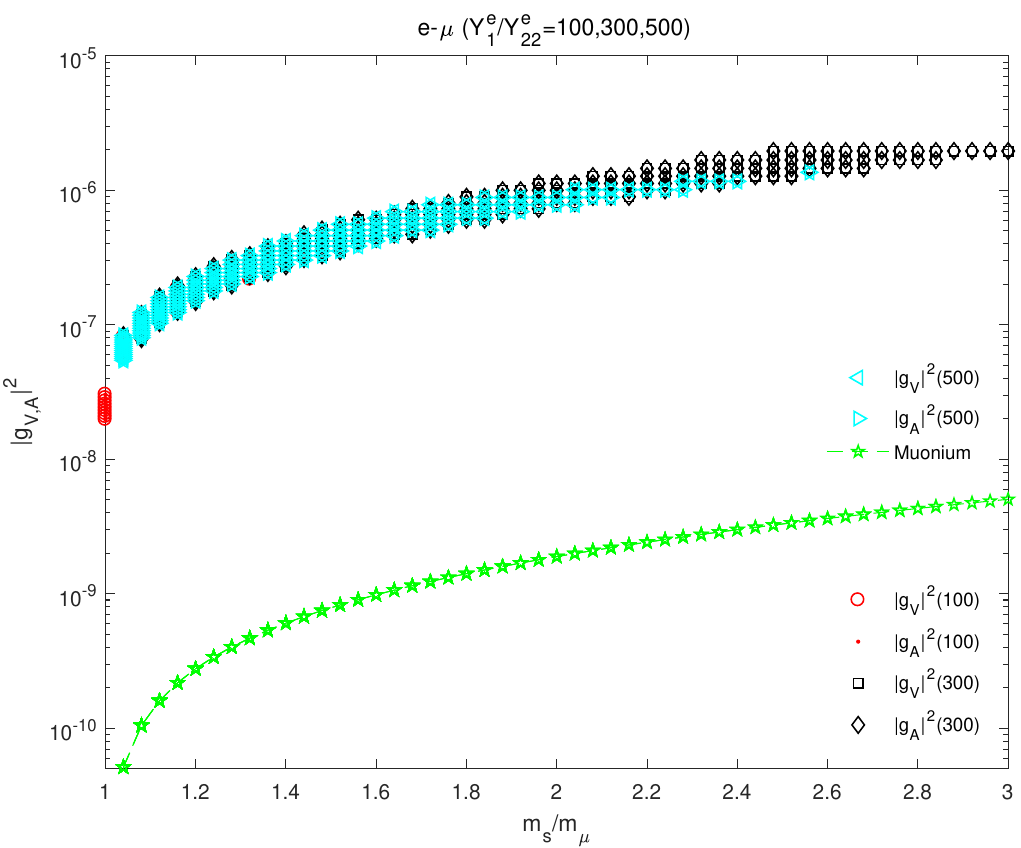}
    }
    {
    \includegraphics[width=0.45\linewidth]{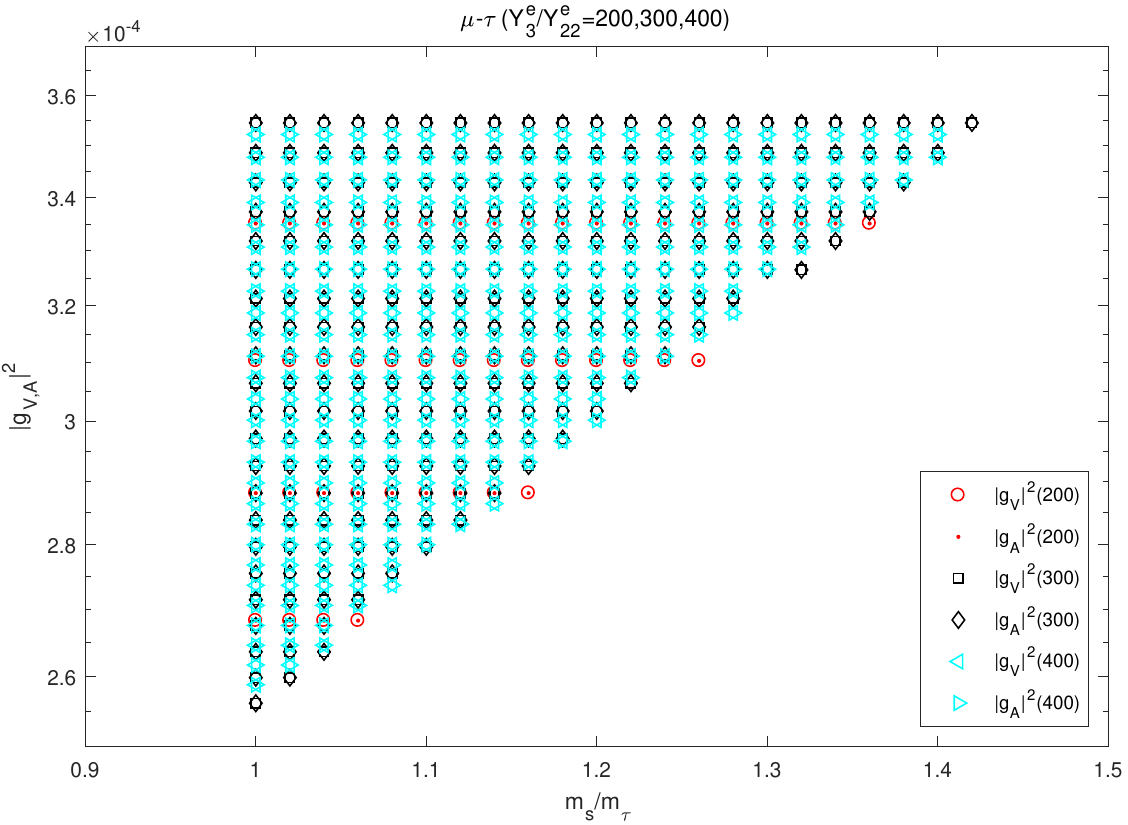}
    }
    \captionsetup{justification=raggedright, singlelinecheck=false}
    \caption{Project the predicted LFV flavon in the mass-coupling plane, corresponding to the $s\bar\mu e$ scenario in Fig.~\ref{fig:abab} (left) and the $s\bar\mu \tau$ scenario in Fig.~\ref{fig:lfmutauflip} (right). The LFV coupling is written as $g_V s\bar\mu e +ig_A s\bar \mu \gamma_5 e+h.c.$, and so on. In the left, the dashed green line represents  the exclusion from the $M-\bar M$ conversion.
    }
    \label{fig:sngva}
\end{figure}

To demonstrate the results, we perform a numerical analysis of the parameter space spanned by $m_L$ and $m_s$, taking three typical ratios $Y^{e}_{1}/Y^{e}_{22}=10,30,50$ for comparison and, moreover, fixing $M^{\ell}_{2}/m_L=10^{-6}$ and $\lambda_2^\ell=1$ except for the last case where $\lambda_2^\ell=0.5$ in order to obtain the required $2\sigma$ confidence muon $g-2$ band (sandwiched between two red lines) in the lighter interval of $m_L$. The results are displayed in the plane $m_L - m_s/m_\mu$ in~\figref{fig:abab}. In these parameter spaces, the constraints from the decay of the $Z$ boson are too weak to yield a meaningful bound. However, muon-electron conversion makes a strong exclusion, leaving the allowed region marked in yellow, whose area is mainly controlled by $M^{\ell}_{2}/m_L$. For a sufficiently large mixing parameter $Y^{e}_{1}/Y^{e}_{22}$, the required value of muon $g-2$ can be accommodated in a wide region, without the need for a high degeneracy between flavon and muon. By the way, one can see that, when $Y^{e}_{1}/Y^{e}_{22}$ decreases, a relatively lighter $m_L$ is necessary to enhance the muon $g-2$, so it is on the frontier of LHC exclusion/discovery. Unfortunately, the muonium-antimuonium conversion rules out the whole parameter space obtained previously~\footnote{It is of interest to investigate if the muonium-antimunium oscillation can be relaxed in some special circumstances; for instance, it receives multiple contributions, and the interference effect may lead to a suppressed oscillation rate. }; it is not shown in these plots, but one can see this in the profile of the flavon obtained in this scenario, the left of Fig.~\ref{fig:sngva}. 

To end up with this part, we note that the same LFV flavon also induces the electron $g-2$ with $\delta a_{e}^{s} \approx \delta a_{e}^{s}(\mu)$ which is strongly correlated with $\delta a_\mu^s(e)$, and they take the ratio
\begin{equation}
   \frac{ \delta a_{e}^{s}(\mu) }{ \delta a_{\mu}^{s}(e) } 
    \approx 
    \f{m_{e}^2}{m_\mu^2}  
    {F_s \left(\frac{m_{\mu}^2}{m_{s}^2}, \frac{m_{e}^2}{m_{s}^2}\right) }/{F_s \left(\frac{m_e^2}{m_{s}^2}, \frac{m_{\mu}^2}{m_{s}^2}\right)},
\end{equation}
When $m_s=m_\mu$, the integral is $ F_s \left(1, {m_{e}^2}/{m_{\mu}^2}\right) = 0.083\ll F_s \left({m_e^2}/{m_{\mu}^2}, 1\right) = 3.85 $. Therefore, $\delta a_{e}^{s}$ is relatively suppressed by the chiral factor $m_e^2/m_\mu^2$ and, moreover, by the loop factor. The model predicts the deviation $\delta a_{e}\sim 10^{-15}$, which is far below the current bound.

\subsection{The last opportunity: \texorpdfstring{$Y_3^e\neq 0$}{Y3e0} and \texorpdfstring{$\tau-\mu-s$}{tau-mu} loop}\label{smutau}

The stringent bound from the muonium and anti-muonium conversion is specifically tailored to the LFV $s\bar\mu e$, and therefore we can still rely on the $s\bar\mu \tau$ scenario; we will elaborate on whether the minimal parameter setting with $Y_3^e\neq 0$ can succeed in this model.

The analysis is similar to the previous case. However, in the current case, the $\mu-f-s$ loops with $f= (\tau, e_{4})$ do not accommodate a light scale. The integral denominator shown in Eq.~\eqref{eq:lfint} is dominated by the mass of the heavy fermion, that is, $z m_s^2 + (1-z) m_{f}^2 - z(1 - z) m_{\mu}^2 \approx (1-z) m_{f}^2$. Therefore, a light flavon with a mass around $m_\tau$ gives a relatively small enhancement. Taking the approximation, $F_s\simeq 1/6m_f^2$ and $G_s\simeq 1/m_f^2$, we estimate 
\begin{equation}
\begin{split}
   \delta a_{\mu}^{s}(f)   \approx  & \f{|\lambda^\ell_2|^2m_{\mu}^2}{32 \pi^2 m_f^2}\times \left[ \frac{1}{6}\times (|(U_L)_{2 2} (U_R)_{4 f}|^2 + |(U_L)_{2 f} (U_R)_{4 2}|^2) 
   \right. \\
   &\left.\hspace{5.5em} + \f{m_f}{m_{\mu}} {\rm Re}\left((U_L)_{2 2}^* (U_R)_{4 f} (U_L)_{2 f}^* (U_R)_{4 2}\right)\right]\,.
\end{split}
\end{equation}
For the heavy fermion contribution, its size $\delta a_{\mu}^{s}(e_4)$ is similar to the previous situation and negligible for the TeV scale VLL.

\begin{figure}
    \centering   
    {
    \includegraphics[width=0.43\linewidth]{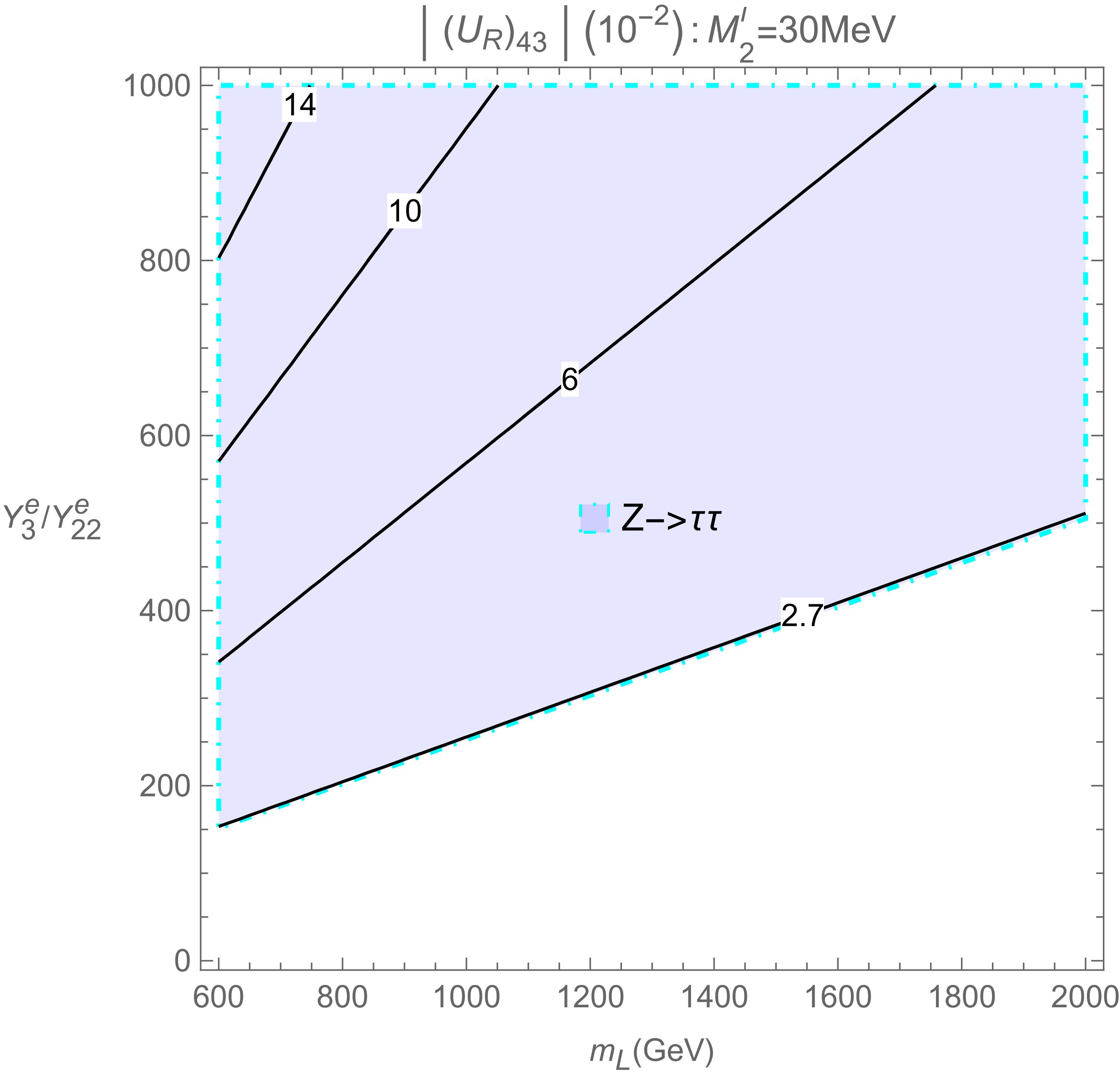}
    }
    {
    \includegraphics[width=0.43\linewidth]{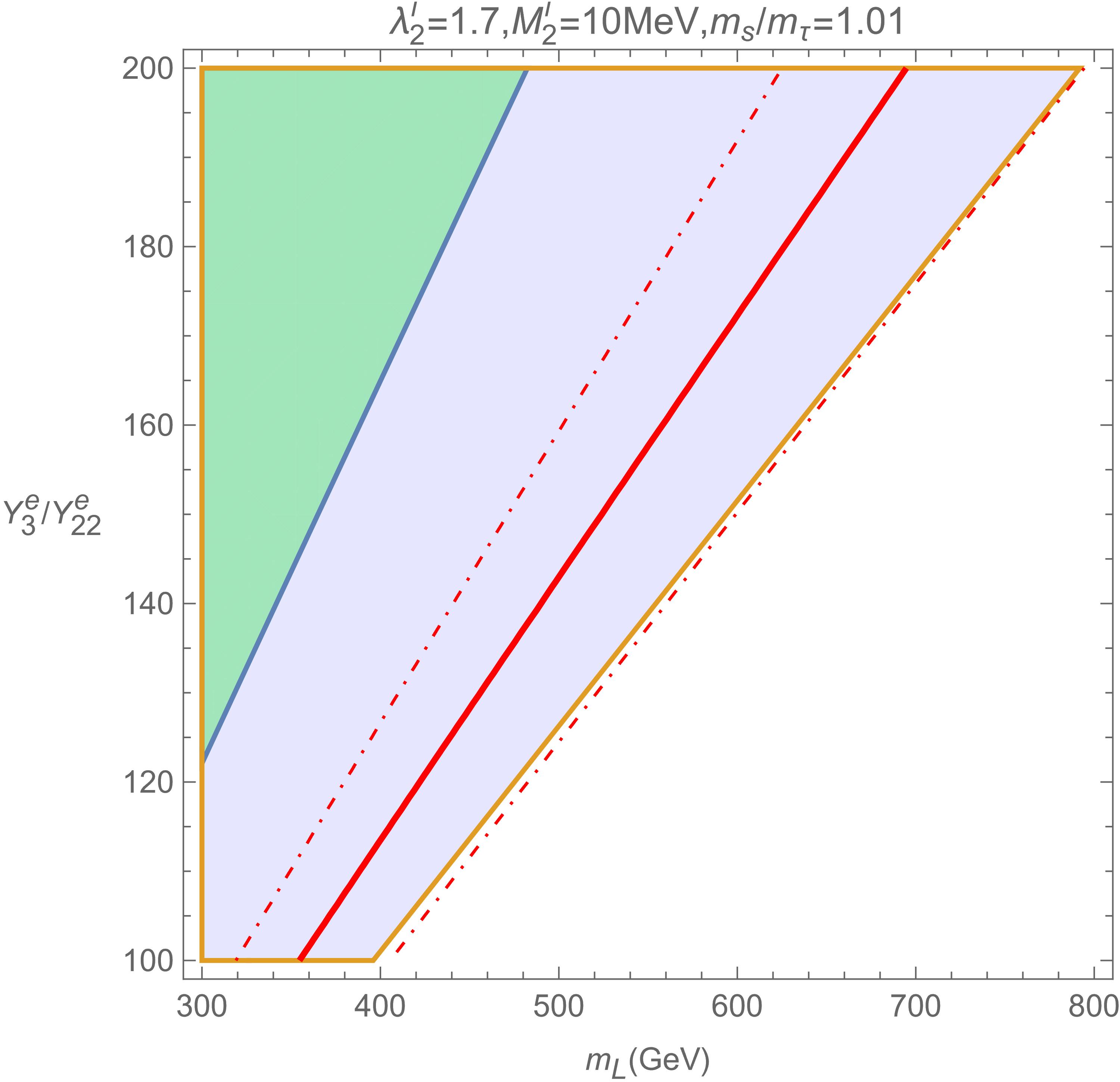}
    }
    \captionsetup{justification=raggedright, singlelinecheck=false}
    \caption{Left: exclusion (blue shaded) from the LFU of $Z$ boson decay in the $s\bar\mu\tau$ scenario, with contours for $|(U_R)_{43}|$ in the unit of $10^{-2}$. Right: demonstration of the solution to the muon $g-2$ puzzle with a minimal $\lambda_2^\ell$, where the band between the dashed lines lives in the edge of LFU exclusion.}
    \label{fig:la}
\end{figure}
Considering the contribution of the $\tau-\mu-s$ loop, it does not benefit from either chiral enhancement nor mitigated loop suppression, see the right panel~\figref{fig:lf2ver}. Therefore, we have to rely on the proper right-handed chiral rotation as large as possible, to make $|(U_L)_{2 2} (U_R)_{4 3}|^2 + |(U_L)_{2 3} (U_R)_{4 2}|^2\sim {\cal O}(10^{-4})$ for $\lambda_a^\ell\sim 1$; a complete parallel analysis to those made in Subsection~\ref{flavon:emu}, it basically requires $(U_R)_{4 3}\simeq \f{b |y_\mu|}{\sqrt{|y_\mu|^2 + |a b|^2}}\gtrsim {\cal O}(0.01)$, with $b=Y_3^e v/(\sqrt{2}m_L)$ here while $a$ remains the same. However, from the subsection~\ref{Zdecay0} it is known that that large element of $U_R$ suffers a stringent constraint from LFU of $Z$ decay,manifest in the $m_L-Y_3^e/Y_{22}^e$ plane in the left of Fig.~\ref{fig:la}, which gives the upper bound on $Y_3^e/Y_{22}^e$. Thus, one needs a moderately large Yukawa coupling $\lambda_2^{\ell}$ to alleviate the burden on $(U_R)_{43}$. Of course, this scenario also gives rise to $\tau$ cLFV decay, but the current experimental bounds are weaker than those of $\mu$ by about four orders of magnitude, see Table~\ref{tab:my_labelmutau}. Nevertheless, note that at the same time, compared to the $\mu-e-s$ loop solution to the muon $g-2$ puzzle, $\Gamma(\tau\to \mu \gamma)$ enjoys a significant enhancement due to the heavier $\tau$ mass (see Eq.\eqref{cLFV:1}), so the cLFV $\tau$ decay yields an even stronger exclusion of the parameter space. Therefore, as in the previous case, we need to substantially suppress $a=M_2^\ell/m_L$ thus $Y_{22}^s$ to reach the surviving parameter space. 

\begin{figure}
    \centering
    {
    \includegraphics[width=0.3\linewidth]{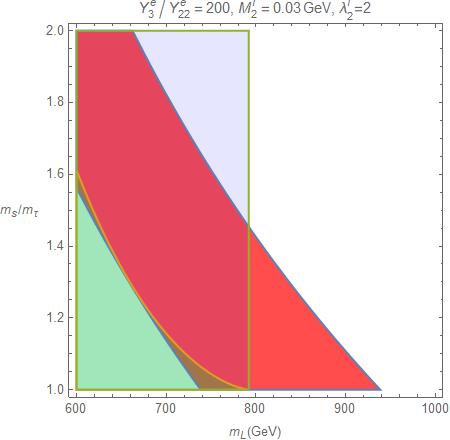}
    }
    {
    \includegraphics[width=0.3\linewidth]{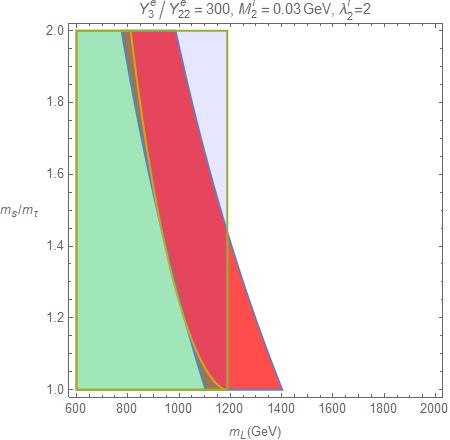}
    }
    {
    \includegraphics[width=0.3\linewidth]{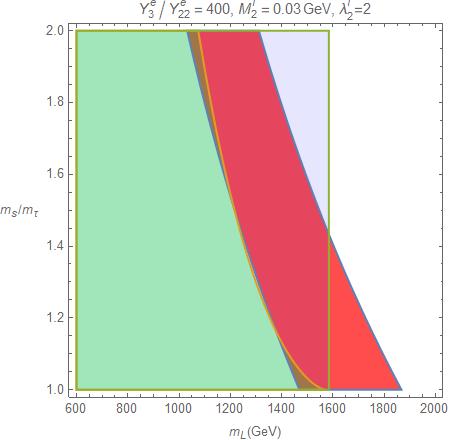}
    }
    \captionsetup{justification=raggedright, singlelinecheck=false}
    \caption{The $\tau-\mu-s$ loop to explain the muon $g-2$ discrepancy  (red-shaded region at the $2\sigma$ confidence level) confronting the LFU from $Z$ decay (blue region) and  cLFV of $\tau\to \mu\gamma$ (green region), taking $M_2^\ell = 0.03\;{\rm GeV}$ and $\lambda_2^\ell=2$ and $Y^{e}_{3}/Y^{e}_{2 2}=200, 300,400$ from left to right.}
    \label{fig:lfmutauflip}
\end{figure}
In the numerical analysis, we have to carefully explore the viable parameter space, and we find that setting $\lambda_2^{\ell}\approx 2$ simultaneously with $Y^{e}_{3}$ as large as possible works. In ~\figref{fig:lfmutauflip}, we display the allowed regions (the red-shaded triangle area) in the $m_L-m_s/m_\tau$ plane for three values of $Y^{e}_{3}$ in units of the muon Yukawa coupling, $Y^{e}_{3}/Y^{e}_{2 2}=200, 300, 400$; the larger $Y^{e}_{3}$ shifts the viable region toward the heavier $m_L$ region. As expected, compared to the $s\bar \mu e$ scenario, here we need $Y_3^e\gg Y_1^e$ to enhance $(U_R)_{43}$, thus compensating for the relative suppression from the heavier $m_\tau$. We tried to find out the minimal $\lambda_2^{\ell}$ and obtained the limit solution with $\lambda_2^{\ell}=1.7$, shown to the right of Fig.~\ref{fig:la}. It is the LFU bound that prevents the even smaller $\lambda_2^{\ell}$.

It is of importance to depict the profile of the resulting flavon for future searches. In the left panel of Fig.~\ref{fig:sngva}, plotted in the $m_s/m_\mu-|g_{V,A}|^2$ plane, we display the profile of $s\bar\mu e$-type flavon, having  a mass $\gtrsim \mathcal{O}(m_\mu)$ and LFV coupling $\sim \mathcal{O}(10^{-3})$, which is able to enhance the muon $g-2$ but is clearly excluded by $M-\overline{M}$ conversion.  Again, we require the flavon mass to lie above $m_\tau$ to avoid a large decay branching ratio of $\tau\to \mu+s$. In the right panel of Fig.~\ref{fig:sngva}, the profile of the $s\bar\mu\tau$-type flavon is shown, and we see that it yields a clear prediction, having mass $m_\tau\lesssim m_s\lesssim 1.5 m_\tau$ and LFV coupling $\sim \mathcal{O}(10^{-2})$. We will comment on the prospect of detecting such a flavon in the part of the discussion.


\section{Conclusion and discussion}\label{dis}

The long-standing muon $g-2$ puzzle has been guiding us to go beyond SM. Maybe, it is tied to another puzzle, the origin of neutrino mass and mixing. In this work, we explore the scenario of a light LFV flavon arising in a class of models with a lepton family symmetry, e.g., the local $(B-L)_{13}$, which is a candidate to realize the minimal seesaw. In addition to the flavon, the full neutrino mixing also needs a pair of VLL, and our discussion expands mainly around these terms $\lambda_2^{\ell} \bar{\ell}_2 L_R \mathcal{F}_{\ell}^* + Y^e_{i} \overline{L}_L H e_{Ri} + m_L \overline{L}_L L_R $. Then, we investigate the possibility of a light LFV flavon to resolve that puzzle, and two scenarios are considered in the spirit of Occam's razor principle to reduce the free parameters (which allows us to develop analytical approximations): 
\begin{enumerate}
    \item The flavon with LFV coupling to $e\mu$ by means of a sizable $Y_1^e$. If there was no muonium-antimuonium oscillation bound, we would have an attractive solution to the muon $g-2$ puzzle since it enjoys the maximal enhancement from a light flavon.         
    \item The flavon with LFV coupling to $\mu\tau$ by means of a sizable $Y_3^e$. In response to the strong constraints of the $\tau$  cLFV decay, in particular from the LFU of $Z$-boson decay which is specific to our model, this scenario leaves us with a flavon with mass $m_\tau\lesssim m_s\lesssim 1.5m_\tau$, without invoking a $\lambda_2^\ell$ considerably larger than 2.
\end{enumerate}
At the LHC, the TeV scale VLL produces same-sign multi-lepton signals through $pp\to e_4^-(\to s\mu^-)+e_4^+(\to s\mu^+)$ followed by the highly boosted $s\to e^\pm \mu^\mp$ and $s\to \tau^\pm \mu^\mp$ in the above two scenarios, respectively, and we notice that a similar feature (lepton jet) has been studied in Ref~\cite{Li:2025luf}. Therefore, they have a very clean signal at the LHC, probably with little background, and can be easily hunted, and we will conduct a specific search in a future publication. Note that if the dominant decay channel of $s$ is into a pair of invisible particles, such as neutrino or dark matter, the signal will become much weaker, and it provides a potential way to rescue the low $m_L$ region.

Detecting the predicted LFV flavon is also of importance. For the above two scenarios we studied, the muon $g-2$ inspired flavons have LFV coupling with strength typically $\sim 10^{-3}$ and $\sim 10^{-2}$, respectively. The former has already been excluded, but such a kind of flavon, not related to muon $g-2$ anomaly, is of interest in a wider sense of LFV at the muon collider such as the projected $\mu$TRISTAN~\cite{Calibbi:2024rcm}. The latter, as a viable solution, is of particular importance. At the Belle II experiment, the light flavon can be produced via the process $e^{+} e^{-} \rightarrow \mu^{\pm} \tau^{\mp} s$. If $s\ra \tau^\pm \mu^\mp$, then the same-sign lepton signal $\mu^\pm\mu^\pm\tau^\mp\tau^\mp$ is almost free of SM background, and a large part of the parameter is testable with the data of $\mathcal{O}(10)~{\rm ab}^{-1}$~\cite{Iguro:2020rby}.



\noindent {\bf{Acknowledgements}}

This work is supported in part by the  National Key Research and Development Program of China Grant No.2020YFC2201504 and the National Science Foundation of China (11775086).

\newpage

\appendix

\section{Higgs LFV coupling}\label{h-LFV}

Within SM, the fermion masses come from a single Higgs VEV $\f{v_h}{\sqrt{2}}\equiv \langle H^0\rangle$ and therefore the mass matrix of the fermions is proportional to their Yukawa coupling matrix, which means that they can be diagonalized at the same time. Consequently, no FCNC is mediated by the SM Higgs boson. However, the presence of a heavy vector-like lepton pair spoils that proportionality, due to the mass terms independent of $v_h$, which are $m_L$, $M_{1,2,3}^\ell$ and also the mass term coming from the flavon VEV, shown in Eq.~(\ref{Me:full}); the following discussions can be easily generalized to other parameter patterns. Then, they give rise to the SM Higgs boson induced LFV. 

After EWSB and flaovn developing VEV, the mass terms of the charged lepton and as well their Yukawa couplings with the SM Higgs boson $h$ are given by
 \begin{align} 
    \mathcal{L}_{Y} &= -(M_e)_{ab}\bar{e}_{La}e_{Rb}-Y^e_{ab}h\bar{e}_{La}e_{Rb}+h.c., 
\end{align}
with the charged lepton mass matrix $M_e$ given by Eq.~(\ref{Me:full}). Transforming to the mass basis, one has 
 \begin{align}
    \mathcal{L}_{Y} 
    &= -({\hat {\cal M}}_e)_{cc}\bar{\hat{e}}_{Lc}\bar{\hat e}_{Rc}-(\hat{Y}^e)_{cd}h\bar{\hat{e}}_{Lc}\bar{\hat e}_{Rd}+h.c.,
\end{align}
with 
\begin{align} 
U^\dagger_LM_e U_R={\hat {\cal M}}_e,\quad U^\dagger_L (v_h Y^e) U_R=v_h{\hat {Y}}^e.
\end{align}
Now, the Yukawa coupling matrix $\hat{Y}^e$ is not diagonal. Subtract the above two equations, we can immediately write $ \hat{Y}^e$ as the sum of two parts,
\begin{equation}\label{Yh}
    \hat{Y}^e ={\rm diag}\L \f{m_e}{v_h},\f{m_\mu}{v_h},\f{m_\tau}{v_h},\f{m_4}{v_h}\R- U_L^\dagger\left(\frac{M_e}{v_h}-Y^e \right) U_R,
\end{equation}
where the first part, for the light species, gives the SM-like couplings, whereas the second part describes the deviations to the SM predictions, rewritten as $U_L^\dagger\frac{M_e(v_h\to 0)}{v_h} U_R$.

\section{Approximately diagonalize the charged lepton mass matrix}\label{sec:mixing information}

In this appendix, we try to develop an approximate way to diagonalize the charged lepton mass matrix under the Occam's razor principle,  Eq.~\eqref{3by3}. Its mass squared matrix $M M^{\dagger}$ and $M^{\dagger} M$ given in \eqref{msq} are related to the left-handed and right-handed rotations, respectively; they both take a seesaw-like hierarchical structure~\cite{seesaw,Grimus:2000vj} with $\mathcal{O}(y_{e})\ll \mathcal{O}(a,b,y_{\mu})\ll1$ and thus may admit approximate diagonalization. 



First, we can obtain $U_L$ from the decomposition of the mass squared matrix  $MM^\dagger/|m_L|^2$ with a good approximation. To that end, let us denote it as the block form
\begin{equation}
 MM^\dagger/|m_L|^2=
 \left(\begin{array}{cc} 
 |y_e|^2 & C \\[0.7ex]
C^\dagger &    \Lambda \\
\end{array}\right) \, ~~ 
 {\rm with}~~
 \Lambda =
\left(
\begin{array}{cc} 
 | y_\mu|^2 + \left| a \right|^2 & a \\[0.7ex]
 a^* &   \left| b \right|^2 + 1 \\
\end{array}
\right), 
\end{equation}
and $C=(0~~y_eb^*)$. The electron can be decomposed through the block chiral rotation
\begin{equation}
\tilde{U}_{L} =
\left(
\begin{array}{cc} 
 \sqrt{1 - B B^{\dagger}} & B \\[0.5ex]
 -B^{\dagger} & \sqrt{1 - B^{\dagger} B} \\
\end{array}
\right) \, ~~{\rm with}~~B=C\Lambda^{-1}=\frac{C}{d}\left(
\begin{array}{cc} 
1+|b|^2 & - a \\[0.5ex]
 - a^*& |y_\mu|^2+|a|^2 \\
\end{array}
\right), \, 
\end{equation}
with $d={\det}\Lambda$. Then, we get electron mass and the mass-squared matrix for the 2-4 block
\begin{align}
    \tilde{m}_1^2 & \approx  \left| m_L \right|^2 \big(\left| y_{e} \right|^2 - \frac{\left| y_{\mu} \right|^2 + \left| a \right|^2}{(\left| y_\mu \right|^2+\left| a b \right|^2)} \left| y_{e} \right|^2 \left| b \right|^2 \big)\; ,\\[0.5ex]
    \tilde{m}_2^2 & \approx  \left| m_L \right|^2
    \left(
    \begin{array}{cc} 
     \left| y_{\mu} \right|^2 + \left| a \right|^2 & a \\[0.5ex]
     a^* & 1 + \left| y_{e} \right|^2 + \frac{\left| y_{\mu} \right|^2 + \left| a \right|^2}{(\left| y_\mu \right|^2+\left| a b \right|^2)} \left| b \right|^2 \left| y_{e} \right|^2 \\
    \end{array}
    \right) \,.\label{eq:ULm2tilde}
\end{align}
$\tilde{m}_2^2$ is also hierarchical and thus it can be approximately diagonalized similarly through
\begin{equation}\label{eq:UL'(24)}
\tilde{U}'_{L} =
\left(
\begin{array}{cc} 
 \sqrt{1 - B' B'^{\dagger}} & B' \\[0.5ex]
 -B'^{\dagger} & \sqrt{1 - B'^{\dagger} B'} \\
\end{array}
\right) \, .
\end{equation}
Therefore, the final left-handed rotation $U_L=\tilde{U}_L\tilde{U}_L'$ has elements
\begin{align}\label{UL33}
(U_L)_{1 1} & \approx 1 - \frac{1}{2}\frac{|a b y_e|^2 (1 + |a|^2)}{(|y_\mu|^2 + |a b|^2)^2},\quad 
(U_L)_{1 2}   \approx \frac{y_e b^* |a| \sqrt{1 + |a|^2}}{(|y_\mu|^2 + |a b|^2)},\quad 
(U_L)_{1 4}   \approx 0\;;\\[1.5ex]\notag
(U_L)_{2 1} & \approx \frac{y_e^* b a}{|y_\mu|^2+|a b|^2},\; 
(U_L)_{2 2}   \approx \left(1 - \frac{|y_e b a|^2 (1 + |a|^2)}{2 (|y_\mu|^2 + |a b|^2)^2}\right) \frac{-{a}/{|a|}}{\sqrt{1+|a|^2}} ,\; 
(U_L)_{2 4}   \approx \frac{a}{\sqrt{1+|a|^2}} \; ;\\[1.5ex]\notag
(U_L)_{4 1} & \approx \frac{- y_e^* b |a|^2}{|y_\mu|^2+|a b|^2},\; 
(U_L)_{4 2}   \approx \left(1 - \frac{|y_e b a|^2 (1 + |a|^2)}{2 (|y_\mu|^2 + |a b|^2)^2}\right) \frac{|a|}{\sqrt{1+|a|^2}},\;
(U_L)_{4 4}  \approx \frac{1}{\sqrt{1+|a|^2}}\;.\notag
\end{align}
The elements related to muon and $e_4$ approximate the corresponding left-handed rotations in the scenario of two generations ($b = 0$). The final mass spectrum is  
\begin{equation}
\begin{cases}
 m_1^2  \approx \frac{v^2}{2} \left| m_L \right|^2 \left| y_{e} \right|^2 \left(1 - \frac{\left| b \right|^2}{\left| y_{\mu} \right|^2+\left| b \right|^2\left| a \right|^2}(\left| y_{\mu} \right|^2 + \left| a \right|^2)\right) \; ,\\[1.5ex]
 m_2^2 \approx \left| m_L \right|^2 \frac{ \left| y_{\mu} \right|^2  + \left| a \right|^2 \left( \left| b \right|^2 + \frac{\left| y_{\mu} \right|^2 + \left| a \right|^2}{(\left| y_\mu \right|^2+\left| a b \right|^2)}  \left| y_{e} \right|^2 \left| b \right|^2 \right) }{ 1+ \left| a \right|^2 } 
 \, ,\\[1.5ex]
 m_4^2 \approx \left| m_{L} \right|^2 (1+ \left| a \right|^2) \, .
\end{cases}
\end{equation}
Therefore, the corrections to the light charged lepton mass from the mixing effect are negligible, due to the smallness of $a$ and $b$.   

However, obtaining $U_R$ from $M^{\dagger} M/m_L^2$ becomes more complicated. This is because after the first block rotation $\tilde{U}_R$, we are left with a remaining matrix $\tilde{\epsilon}$,  
\begin{equation}\label{eq:RH_12}
\frac{M^{\dagger} M}{\left|m_L\right|^2}
= 
\left(\begin{array}{cc|c}
    \left|y_e\right|^2 +\left|b\right|^2 & 0 & b^*  \\[0.5ex]
    0 & \left|y_\mu\right|^2 & y_\mu^* a  \\[0.5ex]\hline
    b & y_\mu a^* & \left|a\right|^2 + 1  \\[0.5ex]
\end{array}\right) \,\to\quad
\tilde{\epsilon} = 
\left| m_L \right|^2
\left(
\begin{array}{cc} 
 \left|y_e\right|^2 + \frac{\left| a b \right|^2}{1 + \left| a \right|^2}  & - \frac{(a b)^* y_\mu}{1 + \left| a \right|^2} \\[0.5ex]
 - \frac{a b y_\mu^*}{1 + \left| a \right|^2} & \frac{\left| y_\mu \right|^2}{1 + \left| a \right|^2} \\[0.5ex]
\end{array}
\right)\,  ,
\end{equation}
which in general lacks a hierarchical structure to be diagonalized with a simple expression. But we have two limits which may give rise to good analytical approximations. One is for $|ab|\ll |y_{\mu}|$, leading to $\tilde{\epsilon}\approx I$ and then 
\begin{equation}
\begin{split}\label{eq:UR33.ablly_mu}
&(U_R)_{1 1}  \approx 1 - \frac{\left| b \right|^2}{2(1 + \left| a \right|^2)^2}\;, \;
(U_R)_{1 2}  \approx  \frac{a^* b^* y_\mu}{2(1 + \left| a \right|^2)^2}\;, \quad\quad\quad
(U_R)_{1 4}  \approx \frac{b^*}{1 + \left| a \right|^2} \;;\\
&(U_R)_{2 1}  \approx - \frac{a b y_\mu^*}{2(1 + \left| a \right|^2)^2}\;, \quad
(U_R)_{2 2}  \approx -1 + \frac{\left| a y_\mu \right|^2}{2(1 + \left| a \right|^2)^2}\;,\quad\
(U_R)_{2 4}  \approx \frac{a y_\mu^*}{1 + \left| a \right|^2} \;;\\
&(U_R)_{4 1}  \approx  - \frac{b}{1 + \left| a \right|^2}\;,  \quad\;
(U_R)_{4 2}  \approx  \frac{a^* y_\mu}{1 + \left| a \right|^2} \;,\quad\quad\quad
(U_R)_{4 4}  \approx 1 - \frac{\left| b \right|^2 + \left| a y_\mu \right|^2}{2 (1 + \left| a \right|^2)^2}\;.
\end{split}
\end{equation}
As expected, the entries $(U_R)_{1 4, 4 1, 2 4 , 4 2}$ are controlled by the $(M^{\dagger}M)_{1 4, 41, 24, 42}$ components.
The other limit is for $|ab| \gg |y_{e}|$, and then the rotation matrix of~Eq.\eqref{eq:RH_12} has the analytical expression 
\begin{equation}
\tilde{U}'_R=
\begin{pmatrix}
   \frac{\left| y_\mu \right|}{\sqrt{\left| a b \right|^2 +\left| y_\mu \right|^2}} & \frac{\left| a b \right|}{\sqrt{\left| a b \right|^2 +\left| y_\mu \right|^2}} & 0 \\[0.8ex]
   \frac{a b y_\mu^*/\left| y_\mu \right|}{\sqrt{\left| a b \right|^2 +\left| y_\mu \right|^2}} & - \frac{a b y_\mu^*/\left| a b \right|}{\sqrt{\left| a b \right|^2 +\left| y_\mu \right|^2}} & 0 \\[0.8ex]
   0 & 0 & 1 \\[0.5ex]
\end{pmatrix}.
\end{equation}
Our analytical analysis made in the text is based on the first limit.

\newpage

\noindent {\bf{Acknowledgements}}

This work is supported in part by the National Science Foundation of China (11775086).

\end{document}